\documentclass[11pt]{article}

\newif\ifnotes
\notesfalse

\usepackage{hyperref}
\usepackage{fullpage}
\usepackage{amssymb}
\usepackage{amsmath}
\usepackage{amsthm}
\usepackage{amsfonts}
\usepackage{bm}
\usepackage{enumitem}
\usepackage{color}
\usepackage{comment}
\usepackage[capitalize]{cleveref}
\usepackage[dvipsnames]{xcolor}
\usepackage{float}
\usepackage[T1]{fontenc}
\usepackage{todonotes}
\usepackage{asymptote}
\usepackage{mdframed}
\usepackage[most]{tcolorbox}
\usepackage{hyperref}
\usepackage{enumitem}
\usepackage{framed}
\usepackage{mdframed}
\usepackage{scrextend}

\usepackage[T1]{fontenc}

\definecolor{denim}{rgb}{0.08, 0.38, 0.74}
\definecolor{periwinkle}{rgb}{0.6, 0.6, 0.95}
\definecolor{wildblueyonder}{rgb}{0.64, 0.68, 0.82}

\usepackage{hyperref}
\hypersetup{
    colorlinks=true,
    linkcolor=denim,
    filecolor=denim,
    citecolor=periwinkle,
    urlcolor=periwinkle,
}
\usepackage[hyperpageref]{backref}

\newtheorem{theorem}{Theorem}[section]
\newtheorem{definition}[theorem]{Definition}
\newtheorem{lemma}[theorem]{Lemma}

\newtheorem{proposition}[theorem]{Proposition}
\newtheorem{corollary}[theorem]{Corollary}

\theoremstyle{remark}

\Crefname{theorem}{Theorem}{Theorems}
\Crefname{claim}{Claim}{Claims}
\Crefname{lemma}{Lemma}{Lemmas}
\Crefname{proposition}{Proposition}{Propositions}
\Crefname{corollary}{Corollary}{Corollaries}
\Crefname{definition}{Definition}{Definitions}

\newcommand{\ECC}{\mathsf{ECC}}
\newcommand{\DEC}{\mathsf{DEC}}

\newcommand{\iECC}{\mathsf{iECC}}
\newcommand{\poly}{\text{poly}}

\newcommand{\Mod}[1]{\ (\text{mod}\ #1)}

\newcommand{\last}{\mathsf{last}}
\newcommand{\mes}{\mathsf{mes}}
\newcommand{\true}{\mathsf{true}}
\newcommand{\false}{\mathsf{false}}
\newcommand{\sent}{\mathsf{cnfm}}
\newcommand{\rec}{\mathsf{rec}}

\newcommand{\cnt}{\mathsf{cnt}}
\newcommand{\knt}{\mathsf{knt}}
\newcommand{\ques}{\mathsf{ques}}
\newcommand{\parity}{\mathsf{par}}

\newcommand{\counter}{\mathsf{counter}}
\newcommand{\stg}{\mathsf{stg2}}
\newcommand{\maj}{\text{maj}}

\newcommand{\bbN}{\mathbb{N}}

\newcommand{\bbZ}{\mathbb{Z}}

\makeatletter
\newcommand{\customlabel}[2]{%
   \protected@write \@auxout {}{\string \newlabel {#1}{{#2}{\thepage}{#2}{#1}{}} }%
   \hypertarget{#1}{#2}
}
\makeatother

\newcommand{\protocol}[3]{
    \stepcounter{figure}
    \vspace{0.15cm}
    { \small
    \begin{tcolorbox}[breakable, enhanced, colback=wildblueyonder!20]
    \begin{center}
    {\bf \underline{Protocol~\customlabel{prot:#2}{\thefigure}: #1}}
    \end{center}
    
    #3
    \end{tcolorbox}
    }
}

\newcounter{casenum}
\newenvironment{caseof}{\setcounter{casenum}{1}}{\vskip.5\baselineskip}
\newcommand{\case}[2]{
    \vskip.5\baselineskip\par\noindent 
    {\bf Case \arabic{casenum}:} {\it #1} \vskip0.1\baselineskip
    \begin{addmargin}[1.5em]{1em}
    #2
    \end{addmargin}
    \addtocounter{casenum}{1}
}

\newcounter{subcasenum}

\newcounter{casenump}
\newenvironment{caseofp}{\setcounter{casenump}{1}}{\vskip.5\baselineskip}
\newcommand{\casep}[2]{
    \ifthenelse{\equal{\value{casenump}}{0}}{
    \vskip.5\baselineskip\par\noindent
    }{}
    {\it Case \arabic{casenump}:} {\it #1}
    \vskip0.1\baselineskip
    \begin{addmargin}[1.5em]{1em}
    #2
    \end{addmargin}
    \addtocounter{casenump}{1}
}

\newcounter{subcasenump}
\newenvironment{subcaseofp}{\setcounter{subcasenump}{1}}{\vskip.5\baselineskip}
\newcommand{\subcasep}[2]{
    \vskip.5\baselineskip\par\noindent 
    {\it Subcase \arabic{casenump}.\arabic{subcasenump}:} {\it #1} \vskip0.1\baselineskip
    \begin{addmargin}[1.5em]{1em}
    #2
    \end{addmargin}
    \addtocounter{subcasenump}{1}
}

\begin{document}

\title{Interactive Error Correcting Codes Over Binary Erasure Channels Resilient to $>\frac12$ Adversarial Corruption}
\author{Meghal Gupta \thanks{E-mail:\texttt{meghal@mit.edu}}\\Microsoft Research \and Yael Tauman Kalai\thanks{E-mail:\texttt{yael@microsoft.com}}\\Microsoft Research and MIT \and Rachel Yun Zhang\thanks{E-mail:\texttt{rachelyz@mit.edu}. Part of this work was done while at Microsoft Research. She is supported by an Akamai Presidential Fellowship.}\\MIT}
\date{\today}

\sloppy
\maketitle
\begin{abstract}
An error correcting code ($\ECC$) allows a sender to send a message to a receiver such that even if a constant fraction of the communicated bits are corrupted, the receiver can still learn the message correctly. Due to their importance and fundamental nature, $\ECC$s have been extensively studied, one of the main goals being to maximize the fraction of errors that the $\ECC$ is resilient to. 

For adversarial erasure errors (over a binary channel) the maximal error resilience of an $\ECC$ is $\frac12$ of the communicated bits. In this work, we break this $\frac12$ barrier by introducing the notion of  an {\em interactive error correcting code} ($\iECC$) and constructing an $\iECC$ that is resilient to adversarial erasure of $\frac35$ of the total communicated bits. We emphasize that the adversary can corrupt both the sending party and the receiving party, and that both parties' rounds contribute to the adversary's budget.

We also prove an impossibility (upper) bound of $\frac23$ on the maximal resilience of any binary $\iECC$ to adversarial erasures. In the bit flip setting, we prove an impossibility bound of $\frac27$. 




\end{abstract}
\thispagestyle{empty}
\newpage
\tableofcontents
\pagenumbering{roman}
\newpage
\pagenumbering{arabic}

\section{Introduction}

Consider the following task: Alice wishes to communicate a message to Bob such that even if a constant fraction of the communicated bits are adversarially tampered with, Bob is still guaranteed to be able to determine her message. This task motivated the prolific study of \emph{error correcting codes}, starting with the seminal works of~\cite{Shannon48,Hamming50}. An error correcting code encodes a message $x$ into a longer codeword $\ECC(x)$, such that the Hamming distance between any two distinct codewords is a constant fraction of the length of the codewords. To communicate a message $x$, Alice sends Bob the corresponding codeword $\ECC(x)$, and the fact that the distance between any two codewords is large guarantees that an adversary must corrupt a large fraction of the communication in order for Bob to decode to the wrong codeword. 

An important question in the study of error correcting codes is determining the maximal possible error resilience.  Indeed, many works focus on precisely this question.  In general, the error resilience parameter depends on the alphabet size: in this work we focus on the binary alphabet.  Two main error models are considered in the literature:  The bit-flip model, where the adversary can flip a constant fraction of the bits of the codeword, and the erasure model, where the adversary can erase a constant fraction of the bits of the codeword (where each erased bit is replaced by a special erasure symbol).\footnote{Another common model that was considered is the insertion/deletion model, which we do not address in this work.}  It is known that in the adversarial bit-flip model, any $\ECC$ can be resilient to at most $\frac14$ corruptions, and in the adversarial erasure error model any $\ECC$ can be resilient to at most $\frac12$ corruptions.

In this work, we investigate the following natural question:

\begin{center}
{\em Can we achieve better error resilience if we use interaction?}
\end{center}

This question is what motivated the study of {\em $\ECC$'s with feedback} \cite{Berlekamp64}, where after every bit sent the sender receives some information about the bits the receiver has received so far.  This feedback is usually noiseless and does not count towards the adversary's budget: error resilience is measured as a fraction of Alice's forward rounds. 
Although noiseless feedback is known to be useful in increasing (forward) error resilience, little is known about noisy feedback in the setting of adversarial corruptions.

In our paper, we show that adversarially noisy feedback can be used to achieve forward erasure resilience $> \frac12$. In fact, we show something even stronger: that adversarially erasable feedback can be used to achieve \emph{total} erasure resilience $> \frac12$ --- for forward and feedback rounds, combined! 
We define the model of {\em interactive error correcting codes} ($\iECC$), where the adversary is given a corruption budget which is an $\alpha$-fraction of the total communication (for some $\alpha>0$). She can spend it arbitrarily (e.g., all on forward communication, or on some combination of forward and feedback communication).\footnote{This is even more general than past considerations of noisy feedback~\cite{WangQC17}, where the adversary is given a separate budget to corrupt forward and feedback messages. We remark that all that was previously known in this setting is that the forward communication can be resilient to $\frac12$ erasures --- which is already achievable with standard (non-interactive) error correcting codes.}
Quite surprisingly, we show that this model allows us to boost the erasure resilience of the protocol beyond $\frac12$. In particular, we show that there is an $\iECC$ for which Bob correctly learns Alice's message $x$ \emph{even if $\frac35$ of the total communication is erased!}

We view $\iECC$'s as the most general and natural solution to the original task of Alice communicating a message $x$ to Bob under adversarial erasure. Note that a standard (non-interactive) $\ECC$ is an $\iECC$ in which Alice speaks in every round. Our result essentially shows that Bob talking occasionally \emph{instead of Alice} actually improves the erasure resilience. It is not obvious that this should be the case! In particular, Bob can only send feedback since he has no input of his own, while Alice can actually send new information. Thus Bob's messages seem less valuable than Alice's. How can the erasure resilience be improved when messages containing information about $x$ have been replaced by messages without? Nevertheless, we show that these feedback messages play a vital role in increasing the erasure resilience past $\frac12$.

As mentioned above, we demonstrate that  $\iECC$'s \emph{are more powerful} than traditional $\ECC$'s by constructing an $\iECC$ that provides better error resilience for adversarial erasure corruptions. But there may be many other potential benefits they may have over traditional $\ECC$'s. Can $\iECC$'s achieve better error rate in other error models (such as the bit-flip model)? 
Do $\iECC$'s have any practical advantages?  We leave these questions to investigation in future work.

\subsection{Our Results.}

Our main result is an $\iECC$ that achieves an erasure resilience of $\frac35$ over the binary erasure channel.
\begin{theorem}
    For any $\epsilon > 0$, there exists an $\iECC$ over the binary erasure channel resilient to $\frac35-\epsilon$ erasures, such that the communication complexity for inputs of size $n$ is $O_\epsilon(n^2)$.
\end{theorem}

We also prove the following impossibility result.

\begin{theorem} \label{thm:intro-2/3}
    There does not exist an $\iECC$ over the binary erasure channel resilient to $\frac23$ erasures.
\end{theorem}

We also consider the problem of constructing an $\iECC$ over the binary \emph{bit flip} channel. In this setting, instead of erasing bits, an adversary may toggle bits of her choice. 
We do not know of any constructions of $\iECC$s that achieve a higher error resilience to bit flips than the standard (non-interactive) error correcting code (which achieves a maximal error rate of $\frac14$). We leave open the problem of constructing such an $\iECC$.
We prove the following negative result on the maximal possible error resilience to bit flips of any $\iECC$ scheme.

\begin{theorem} \label{thm:intro-2/7}
    There does not exist an $\iECC$ over the binary bit flip channel resilient to $\frac27$ adversarial bit flips.
\end{theorem}

\subsection{Related Work}

We mention two areas of research most pertinent to our work. The first is  {\em error correcting codes with feedback} and the second is {\em interactive coding}.

\subsubsection{Error Correcting Codes with Feedback}\label{sec:related:feedback}

The notion of an \emph{error correcting code with feedback} was first introduced in the PhD thesis of Berlekamp~\cite{Berlekamp64}. In an error correcting code with feedback, Alice wishes to communicate a message to Bob in an error resilient fashion. For every message she sends, Bob sends her feedback. Originally, this feedback was considered in the \emph{noiseless} setting, meaning that none of Bob's messages are allowed to be corrupted. Furthermore, Bob's feedback is \emph{not} counted towards the adversary's corruption budget. That is, error rate is calculated solely as a function of the number of messages Alice sends.


In the bit flip error model, \cite{Berlekamp68,Zigangirov76,SpencerW92,HaeuplerKV15} showed that the maximal error resilience of an error correcting code with noiseless feedback is $\frac13$. For larger alphabets, the maximal error resilience was studied in~\cite{AhlswedeDL06}. In the erasure model, error correcting codes with feedback over any alphabet are known to be resilient to an arbitrarily close to 1 fraction of erasures.\footnote{While we don't know of an explicit reference for this, this can be seen using ideas from this paper or from~\cite{GuptaZ21a}. Essentially Alice can send her input bit by bit, only moving onto the next bit when she receives confirmation that Bob has received the last bit.} 

When the feedback is \emph{noisy}, i.e. the feedback may be corrupted as well, much less is known. Several works such as \cite{BurnashevY08a,BurnashevY08b} considered $\ECC$'s with noisy feedback over the binary symmetric channel (each communicated bit is independently flipped with some probability). In the case of adversarial corruptions, the only work we know is that of~\cite{WangQC17}, which places separate corruption budgets on the forward and feedback rounds. They construct a scheme that is resilient to $\frac12$ of the forward communication and $1$ of the feedback being erased.\footnote{Our scheme achieves this as well.} We note that their scheme's forward erasure resilience is equal to that achievable by standard error correcting codes. 

As far as we know, our notion of an interactive error correcting code, where the adversary is permitted to freely corrupt both Alice's bits and Bob's feedback up to a constant fraction of the total communication, has not been studied before.

\subsubsection{Interactive Coding}\label{sec:IC}

Interaction has been considered in the context of error resilience starting with the seminal works of Schulman~\cite{Schulman92,Schulman93,Schulman96} and continuing in a prolific sequence of followup works, including~\cite{BravermanR11,Braverman12,BrakerskiK12,BrakerskiN13,Haeupler14,BravermanE14,GellesHKRW16, GellesH17,EfremenkoGH16,EfremenkoKS20b}. These works consider the task of taking an {\em interactive} protocol and making it error resilient. That is, they consider \emph{two-way} communication, instead of one-way as we do.


More specifically, Alice has a private input $x$, Bob has a private input $y$, and they both want to compute $f(x,y)$. Given a protocol $\pi_0$ that achieves this, but is not necessarily resilient to any error, the goal is to construct a (fixed order, fixed length) 
protocol $\pi$ that is resilient to a constant fraction of adversarial bit flip errors. 
The question of maximizing $\pi$'s error resilience has been studied in~\cite{BravermanR11,GhaffariH13}, albeit in the context of large alphabets.~\cite{BravermanR11} presented the first error-resilient protocol over the binary channel, achieving an error resilience of $\frac18$. This was later improved to $\frac5{39}$ in~\cite{EfremenkoKS20b}, and finally to $\frac{1}{6}$ in the work of~\cite{GuptaZ21a}. $\frac16$ is known to be optimal.

This problem was also considered in the case of adversarial erasures instead of adversarial bit flip errors~\cite{FranklinGOS15,EfremenkoGH16,GellesH17,GuptaZ21a}. The work of~\cite{FranklinGOS15} originally constructed such a scheme over a large alphabet 
with erasure resilience $\frac12$, which translates to a resilience of $\frac14$ over the binary erasure channel by encoding each original letter with a binary error correcting code. The work of~\cite{EfremenkoGH16} improved the binary erasure resilience to $\frac13$, and the recent work of~\cite{GuptaZ21a} settled this question, showing that the optimal erasure resilience is $\frac12$, which is known to be optimal~\cite{FranklinGOS15}.

We note that there are many other related works, some which consider adaptive speaking order and adaptive length protocols, where the adaptivity is added in an effort to improve the error resilience. There are also works on interactive coding in the multi-party setting.  We refer the reader to \cite{Gelles-survey} for details.

\section{Technical Overview} \label{sec:overview}

In this overview, we focus our efforts on our constructions of an $\iECC$ which bypass the $\frac12$ error resilience barrier.  We start with presenting a simplified construction that achieves erasure resilience of $\frac6{11}$, and then present our improved (complicated) construction that achieves erasure resilience of $\frac35$. For the reader simply interested in how to construct $\iECC$'s better than traditional $\ECC$'s, it is sufficient to understand the $\frac6{11}$ protocol. For the impossibility bounds given in Theorems~\ref{thm:intro-2/3} and~\ref{thm:intro-2/7}, we refer the reader to Sections~\ref{sec:2/3} and~\ref{sec:2/7} respectively. 

\subsection{A Simpler $6/11$ $\iECC$}\label{sec:6/11:overview}

We begin by discussing a simpler scheme that achieves an error resilience over the binary erasure channel of $\frac6{11} - \epsilon$.\footnote{For simplicity, we omit $\epsilon$'s for the rest of this overview; all fractions $r$ should be understood to be $r \pm O(\epsilon)$.} Our starting point is based on the list-decoding to unique-decoding paradigm of~\cite{GhaffariH13,EfremenkoKS20b}.
\begin{enumerate}
    \item Alice sends Bob $\ECC(x)$, where $\ECC$ is an error correcting code of distance $\frac12$.  
    
    Denote by $M$ the length of the codeword $\ECC(x)$. As we show in Lemma~\ref{lem:1/4}, as long as less than $\frac34$ of the bits are erased, there are at most two codewords that agree with the unerased bits of $\ECC(x)$. Furthermore, since the adversary can only erase and not flip bits, we have a perfect guarantee that one of these codewords is $\ECC(x)$.
    
    \item Bob sends Alice an index $i$ on which the two inputs differ, using an error correcting code.
    \item Alice decodes Bob's message, and replies with the value of her input on index $i$ (which is sufficient for Bob to deduce $x$). She can do this by simply sending $0^M$ or $1^M$. Bob now only needs to hear any \emph{one} of Alice's message to learn her input.
\end{enumerate}

Individually, both steps 1 and 3 are resilient to $\frac34$ erasures, which gives hope for ultimately constructing a protocol resilient to $>\frac12$ erasures.  Unfortunately, there are some glaring issues.  First, the fact that each message is resilient to $>\frac12$ erasures does not imply that the final protocol is resilient to $>\frac12$ erasures, since the adversary can choose to divide his corruption budget arbitrarily, and in particular can corrupt much more of one message at the cost of corrupting less of another. Furthermore, there is no obvious way for Bob to communicate $i$ to Alice in a way resilient to $>\frac12$ erasures. If Bob sends $\ECC(i)$, an adversary can simply erase half of Bob's message, making this step only $\frac12$ error resilient. 

This latter problem is addressed as follows: instead of having Bob send $\ECC(i)$, we limit Bob's message space to consist of only four possible messages, which can have relative distance $\frac23$ (e.g. $\bar{0}=(000)^*, \bar{1}=(011)^*, \bar{2}=(101)^*, \bar{3}=(110)^*$). At this point the reader should ask:  {\em How can Bob communicate the index $i$ to Alice while only sending one of four possible messages?} To do this, we must use interaction. In our scheme, Bob communicates $i$ to Alice via an \emph{incrementation procedure} consisting of many rounds of interaction in which Bob always sends one of two codewords $\bar{0}, \bar{1}$. The other two codewords will be used to communicate some additional information that we will specify later.

\paragraph{The incrementation procedure.}
In our incrementation procedure, Alice keeps track of a counter $\cnt$ initially set to $0$ indicating her guess for $i$. In each round, she sends $x$ and $\cnt$, jointly encoded with a distance-$\frac12$ error correcting code, to Bob. Bob's goal is to increment $\cnt$ to $i$ by sending just two codewords $\bar{0}, \bar{1}$. 

As a first attempt, one could consider a scheme in which Alice increments $\cnt$ every time she hears $\bar{1}$ from Bob, and stops incrementing when she hears  $\bar{0}$ from Bob. However, there's a clear problem: what should Bob send if he doesn't hear Alice? He doesn't know if she has incremented enough yet, in which case he should send $\bar{0}$, or if she should increment again, in which case he should send $\bar{1}$. If he sends $\bar{1}$ every time he isn't sure, Alice might not know if Bob has heard her last message and wants her to keep incrementing or not, so she might increment past $i$. If he sends $\bar{0}$, the adversary could employ the following attack: she erases the $\bar{1}$ but not the following $\bar{0}$ from Bob, so that she only erases $\frac12$ of Bob's messages (recall we need this to be $>\frac12$), while keeping Alice from incrementing at all.

Instead, we use the following procedure:  Alice increments only when she detects a \emph{change} in Bob's message from $\bar{0}$ to $\bar{1}$ or vice versa. This change in Bob's message signals to her that Bob has heard her latest value of $\cnt$ and wants her to increment again; otherwise, he may not yet know if she's incremented or not. Meanwhile, Bob sends the same message $\bar{0}$ or $\bar{1}$ until he detects that Alice has incremented, before switching to sending the other codeword to ask Alice to increment again. The idea is that Alice will only increment again when Bob has acknowledged her previous incrementation and asked her to increment again, so that the two can never get out of sync. In particular, Alice cannot skip over the index $i$ without Bob's permission. 

\paragraph{Our protocol.}
Our protocol consists of many (say $\approx \frac{n}{\epsilon}$) \emph{chunks}, where in each chunk Alice sends a message followed by Bob's reply.
Our protocol is designed so that each such chunk will make \emph{progress} towards Bob's unambiguously learning Alice's input, as long as the adversary did not invest more than $\frac6{11}$ error in that chunk. At a high level, in the first chunk with $< \frac6{11}$ erasures, Bob narrows down Alice's input to at most two options. In every future chunk with $<\frac6{11}$ erasures, either Alice gets closer to learning the index $i$ on which the two options differ, or Bob fully determines $x$ by ruling out one of the two values of $x$, e.g. by learning the value of $x[i]$ or by uniquely decoding Alice's message.



We choose the parameters so that in each chunk Alice sends a message of length $M$ and Bob replies with a message of length $\frac38M$.  This choice implies that in a chunk with $<\frac6{11}$ erasures, it is guaranteed that either Bob hears $>\frac12$ of Alice's message and thus can uniquely decode it, or he hears $>\frac14$ of Alice's message \emph{and} Alice hears $>\frac13$ of Bob's message (this follows from the $8:3$ message length ratios of Alice and Bob). Therefore, in a chunk with $<\frac{6}{11}$ erasures, it is guaranteed that either Bob uniquely decodes Alice’s message, or Bob narrows down Alice’s message to two options and Alice uniquely decodes Bob’s message (because Bob sends one of four codewords with relative distance $\frac23$). 

Let us describe the protocol. Alice keeps track of a counter $\cnt$ initially set to $0$ indicating her guess for $i$. At the beginning of the protocol, Alice sends $\ECC(x, \cnt)$ to Bob in every chunk. At some point there will be $<\frac6{11}$ erasures in a chunk, and so
Bob list decodes Alice's message to at most two options, say $(x_0, \cnt_0 = 0)$ and $(x_1, \cnt_1 = 0)$. Since we are in the setting of erasures, one of the two decodings must be Alice's true state, and in particular must contain Alice's true input. Note that if instead Bob uniquely decodes Alice's message, he can unambiguously determine her input $x$. In general, the case where Bob uniquely decodes Alice's message allows him to trivially determine $x$, so we do not mention it, and assume that in all chunks with $<\frac6{11}$ error, Bob list-decodes Alice's message to two options, and Alice uniquely decodes Bob's message.

At this point, Bob begins signaling to Alice to increment $\cnt$. His goal is to tell Alice to increment $\cnt$ until $\cnt = i$. To do this, Alice increments $\cnt$ when she sees Bob's messages \emph{change} from $\bar{0}$ to $\bar{1}$ or vice versa. Bob correspondingly waits until he next list-decodes Alice's message to two options, and sees that $\cnt$ has been incremented correctly before flipping his message. Because there are at least $i$ chunks with $<\frac6{11}$ erasures, they will progress $i$ times, and Alice will reach the index $i$. 

Note the counters $\cnt_0$ and $\cnt_1$, obtained by list decoding Alice's message, may not be equal! In this case Bob will learn the correct $x$ in a different way, as we  explain later. For now assume that the two decoded messages are $(x_0,\cnt)$ and $(x_1,\cnt)$.

When Bob sees that Alice's counter has reached the index $i$, he begins sending a third codeword $\bar{2}$ to ask Alice for the value of her input at index $i$. Upon uniquely decoding $\bar{2}$, Alice knows that the index $i$ has been reached and begins sending $x[i]$ for the rest of the protocol. As long as Bob eventually receives one bit of Alice's messages after this point, he can correctly deduce Alice's input~$x$.

Recall, however, that at some point in the protocol the two messages that Bob decodes may have different counters $\cnt_0\neq\cnt_1$. In order to learn $x$, it suffices for Bob to determine Alice's true value of $\cnt$ since $x_b$ and $\cnt_b$ are paired up. If the first time the counters get out of sync one is at least $2$ greater than the other, then Bob can conclude that one made an impossible increment and thus deduce $x$. Otherwise, the counters differ by~$1$, and Bob begins sending a fourth codeword $\bar{3}$, asking Alice to tell him the parity of her counter. This lets Bob deduce which of $\cnt_0$ and $\cnt_1$ was Alice's true counter, and then the corresponding $x$ must be Alice's true input.

At the end of the protocol, Bob takes the last bit he ever received as the answer to his question (value of $x[i]$, corresponding to $\bar{2}$, or parity of $\cnt$, corresponding to $\bar{3}$), and deduces Alice's input $x$.  We refer the reader to \cref{sec:6/11} for a formal description of the protocol and its analysis.

\subsection{Improving $6/11$ to $3/5$}\label{sec:3/5-overview}

In the above protocol, Bob uses the codewords $\bar{0}$ and $\bar{1}$ to increment Alice's counter. Then, when he wants to ask Alice for either the value of $x[\cnt]$ or the parity of $\cnt$, he has to use two \emph{new} codewords $\bar{2}$ or $\bar{3}$ to convey the appropriate question.  By including these two extra codewords, the maximal possible distance between Bob's possible codewords decreases from $1$ to $\frac23$, which incurs a loss in error resilience. 
If we were able to somehow have Bob use only two codewords, the relative distance between Bob's codewords would increase to $1$. Then, by letting Alice and Bob speak in a $4:1$ ratio, to prevent progress an adversary would have to either corrupt $\frac34$ of Alice's message, or all of Bob's and $\frac12$ of Alice's. This increases the erasure resilience to
\[
    \frac{\frac34 \cdot 4M}{4M + M} = \frac{M + \frac12 \cdot 4M}{4M + M} = \frac35.
\]

In order to get rid of the need for the extra two codewords $\bar{2}$ and $\bar{3}$, we need for the two codewords $\bar{0}$ and $\bar{1}$ to be able to take on \emph{more than two meanings}. To do this, we combine our two codewords with \emph{timing cues}, such that $\bar{0}$ and $\bar{1}$ mean different things depending on where they are heard in the protocol. More specifically, we divide the protocol into \emph{blocks} of many messages, such that Alice interprets messages differently for the rest of the block depending on whether she first decoded a $\bar{0}$ or $\bar{1}$ within the block. The bits communicated in the rest of the block can now be combined with the first heard bit to take on more than two meanings. 

There are two pieces of information that Bob needs to convey, which he previously used the extra codewords for. These are:  
\begin{enumerate}
    \item Telling Alice she is done incrementing $\cnt$ (and thus can switch to answering Bob's question for the rest of the protocol). 
    \item Telling Alice which question (value of $x[i]$ or parity  of $\cnt$) to answer.
\end{enumerate}

First, we describe a new incrementation procedure, which ultimately allows Bob to tell Alice to stop incrementing the counter without introducing a new codeword. In this incrementation process, Alice always expects to hear $\bar{1}$ before $\bar{0}$ in every block, so she can reserve hearing $\bar{0}$ as the first message she hears in the block to mean that the incrementation is over. Then, we describe a second set of modifications to this protocol for Bob to specify his question without introducing a new codeword.

\paragraph{A new incrementation procedure.}

We partition the rounds into blocks, each consisting of {\em several} ($\approx \frac1\epsilon$) 
messages. The first message that Alice hears from Bob in each block indicates to her whether to increment her counter or terminate the incrementation stage: namely, Bob sends $\bar{1}$'s if he wants Alice to increment her counter and sends $\bar{0}$'s if he wants her to terminate the incrementation stage. However, if Bob just sends $\bar{1}$'s for the entire block when he wishes for Alice to increment (and $\bar{0}$'s when he wants her to terminate), we run into the same problem as we discussed earlier: Alice's messages may be erased, in which case Bob doesn't know whether Alice has incremented or not. To keep Alice and Bob in sync, Bob sends \emph{confirmation} messages to tell Alice that he saw her previous incrementation and that she should increment again the next time she hears a $\bar{1}$.


To be precise, in each block, Bob attempts to increment Alice's $\cnt$ by exactly $1$. When Bob wishes for Alice to increment her counter, he begins a block by sending $\bar{1}$'s to ask Alice to increment $\cnt$, then when he sees that Alice has incremented $\cnt$, he sends her $\bar{0}$'s for the rest of the block to \emph{confirm} the incremented value of $\cnt$. Alice only increments $\cnt$ again in the next block if her current value of $\cnt$ has already been confirmed. 

To record Bob's confirmations, Alice has another variable $\sent$, in addition to $\cnt$, taking values in $\{ \true, \false \}$. Each time Alice increments $\cnt$ she immediately sets $\sent \gets \false$ until she gets a confirmation (a $\bar{0}$ after receiving at least one $\bar{1}$ within the same block) from Bob, at which point she sets $\sent \gets \true$. Only when $\sent = \true$ does Alice increment $\cnt$ again when she receives a $\bar{1}$ from Bob. 

Meanwhile, when Bob wishes to continue the incrementation stage he sends $\bar{1}$ in a block until he is \emph{sure} that Alice heard him. 
Note that it is not always apparent from Alice's message whether she's heard him within this block or not, since if $\cnt$ is not confirmed she sends the same message $\ECC(x, \cnt, \false)$ whether she heard a $\bar{1}$ this block or not. Therefore, to ensure that Bob receives feedback on whether Alice has heard a $\bar{1}$ this block, Alice adds yet another variable $\rec \in \{ \true, \false \}$ indicating to Bob whether she's received a $\bar{1}$ from him this block. 

In detail, our incrementation procedure is as follows:



\begin{itemize}
    \item 
        Every chunk, Alice sends her input $x$, along with a counter $\cnt$, a boolean $\sent$ detailing if her current value of $\cnt$ is confirmed, and $\rec$ which is $\true$ if she has received a $\bar{1}$ so far this block, all jointly encoded with a standard error correcting code.
    \item 
        Bob begins each block by sending $\bar{1}$'s.
    \item 
        If the first message that Alice receives in a block is $\bar{1}$, she increments $\cnt$ and sets $\sent = \false$ if the previous value of $\cnt$ had been confirmed ($\sent = \true$); otherwise, she does nothing.
    \item 
        If Bob learns that Alice has received a $\bar{1}$ this block, he attempts to confirm her value of $\cnt$ by sending $\bar{0}$ for the rest of the block. 
    
    \item 
        If Alice gets Bob's $\bar{0}$ after having received a $\bar{1}$ in the same block, she confirms her new value of counter by setting $\sent = \true$. She is now ready to increment again the next time she hears a $\bar{1}$.
        
        \item If the first message that Alice receives within a block is a $\bar{0}$, she moves on from this incrementation stage (which we later refer to as Stage 1) to a new stage, either Stage 2 or 3, which we will define and discuss later.  Thus, when Bob wants Alice to terminate incrementation and move on (the equivalent of the old codewords $\bar{2}$ or $\bar{3}$) he sends $\bar{0}$'s the entire block.
\end{itemize}


We want to point to a technical issue that will complicate our protocol.  As mentioned earlier, the adversary has the budget to erase $\frac12$ of all of Alice's messages, and as a result confuse Bob between two Alices. 
What does Bob do when he receives a message from two possible Alices, one with $\rec=\true$ and the other with $\rec=\false$?  Since the protocol must make progress in this case, we instruct Bob to send a confirmation (of the form $\bar{0}$) even if only one of these Alice's has $\rec=\true$.

The above choice can result in the following tricky situation: The adversary can erase Bob's first messages in a block, so that the ``real'' Alice does not receive any $\bar{1}$'s, and at the same time confuse Bob between two Alices, the real which has $\rec=\false$ and a fake which has $\rec=\true$, in which case, Bob will proceed to send confirmation of the form~$\bar{0}$ for the rest of the block. The adversary will not erase these $\bar{0}$'s, and as a result the first message that the real Alice receives in the block is a $\bar{0}$, which will cause her to leave the incrementation stage of the protocol, even though $\cnt$ can be very far from~$i$ (and the counters of the two different Alice's equal). Recall that this situation could not occur in our $\frac6{11}$ protocol (described in Section~\ref{sec:6/11:overview}), since we did not use the same codeword $\bar{0}$ to mean two different things!

To deal with this, Bob's goal will be more generally to guide the two Alices to send \emph{different bits} by the end of the protocol, so that if he hears any such bit he can determine the real Alice. This is achieved via the question-asking paradigm, as we discuss next, but in certain edge cases, he uses different techniques as we discuss later.

\paragraph{Specifying the question.}
Recall that when the first message that Alice receives in a block is $\bar{0}$, she advances to Stage~$2$ or $3$. Intuitively, the purpose of Stage $2$ is for Alice to learn Bob's question (parity or value), and the purpose of Stage $3$ is for Bob to learn Alice's answer. That is, when Alice is in Stage 3, she sends messages of the form $b^{4M}$ where the bit $b$ conveys her answer to Bob's question.

Alice learns Bob's question in Stage 2 similar to the way she learned the index~$i$ in Stage 1.  Namely, she keeps yet another counter denoted by $\knt$, whose purpose is similar to that of $\cnt$ in Stage 1.  In the beginning of stage 2 $\knt$ is initialized to $\knt= 0$. She and Bob then participate in an incrementation procedure (similar to that in Stage 1), where the goal is to keep $\knt=0$ if Bob wishes to learn the bit value $x[i]$, and increment it to $1$ if he wishes to learn the parity of $\cnt$. 

We note that as opposed to the blocks which are of fixed length, and hence the parties always agree on when a block begins and when it ends, the length of each stage is not fixed and may depend on Alice's input and the adversarial corruptions. As a result, Bob may not know which stage Alice is in. 
In an effort to remove this ambiguity,  we double the signal of $\knt$ to also include which stage Alice is in. When Alice is in Stage 1 she sets $\knt=-1$, and when she is in Stages 1 and 2, she sends $\ECC(x, \cnt, \sent, \rec, \knt)$ to Bob, so that if Bob receives a message with $\knt = -1$, he knows she is in Stage 1, and if $\knt \in \{ 0, 1 \}$, he knows she's in Stage 2. When Alice is in Stage 3, her message is of the form $b^{4M}$ for some $b \in \{ 0, 1 \}$, which is also distinguishable.

\paragraph{Dealing with two different Alices.} 
Unfortunately, the same problem of the adversary confusing Bob between two Alices continues to haunt us! Our tools from the $\frac6{11}$ protocol are sufficient only to deal with the case that both Alices are still in Stage 1, the incrementation stage. When this is not the case, e.g. the two Alices are in different stages, we must guarantee that they end up in Stage 3 with opposite bits.

To solve this, we first notice that in many cases Alice can skip Stage~$2$ altogether.  For example, if the bit value $x[i]$ is equal to the parity of $\cnt$, both of which are equal to $0$, she can skip Stage 2 (since $x[i] = \cnt~\text{mod}~2 = 0$ is the correct answer to both questions).  Moreover, we slightly change the incrementation stage so that she can also skip Stage 2 in the case where the parity of $\cnt$ is $1$.  To this end, in the incrementation stage, Bob asks Alice to increment $\cnt$ until $\cnt = 2i$, in which case Alice knows he is interested in $x[\cnt/2]$ or in the parity of $\cnt$. 
Thus, if $\cnt$ is odd there is no associated value question, so Alice knows to send $1$, the parity of $\cnt$, for the rest of the protocol. 


With this change to the protocol, the only case where Alice actually needs to learn Bob's question is when her \emph{parity} bit is $0$ and her \emph{value} bit is $1$. In this case, Alice will advance from Stage~$1$ to Stage 2 by setting $\knt=0$, and in all other cases Alice advances from Stage~$1$ directly to Stage~$3$ and sends her answer for the remainder of the protocol.

Now, lets go back to the question:  {\em What does Bob do if the adversary confuses him between two Alices?} Bob deals with this differently, based on which stages the two Alices are in. 

\begin{enumerate}

\item {\bf Both Alices are in Stage 1.}  
    We've already discussed this case: If either $\cnt_0 = \cnt_1 = 2i$ or $\cnt_0 \not= \cnt_1$, Bob sends $\bar{0}$ for the rest of the protocol. If $\rec_0 \not= \rec_1$, Bob behaves as if Alice heard his message (which unfortunately may cause the Alice with $\rec = \false$ to prematurely exit Stage~$1$).

\item {\bf One Alice is in Stage 1 and the other is in Stage 2.} 
    This is the difficult case!   If Bob tries to increment both counters simultaneously and then advance both simultaneously, as he would if both were in Stage 1, he cannot necessarily coordinate them to send opposite bits for the rest of the protocol, since (as one of many issues) Stage 2 Alice may have left Stage 1 prematurely. We solve this problem by introducing a final layer of grouping, called the \emph{megablock}.

    Our final protocol consists of several ($\approx \frac1\epsilon$) \emph{megablocks}, each containing many ($\approx \frac{n}\epsilon$) blocks. At the beginning of each megablock, Alice resets the counter she is currently incrementing (either $\cnt$ or $\knt$) to $0$. 
    If Bob ever sees two Alices, one of whom is in Stage 1 and the other who is in Stage 2, he waits until the start of the next megablock (meanwhile sending $\bar{1}$'s) and then sends $\bar{0}$ for the rest of the protocol. Recall that the Alice in Stage 2 must have $\cnt ~\text{mod}~ 2 = 0$ and $x[\cnt/2] = 1$, so when she receives a $\bar{0}$, she learns Bob's question to be \emph{value} and sends $1$ for the rest of the protocol. As for the Stage 1 Alice, by convention, we say that if Alice is in Stage 1 and receives a $\bar{0}$ first within a block while her value of $\cnt$ is $0$, she advances directly to Stage 3 and sends $0$ for the rest of the protocol. Thus, when Bob sends $\bar{0}$ for the rest of the protocol starting at the beginning of a megablock, the Stage 1 Alice and the Stage 2 Alice both eventually advance to Stage 3 with opposite bits.
    

    
\item {\bf Both Alices are in Stage 2.}   
    We argue that this will never happen! (Unless it is easy for Bob to detect that one of the Alice's is fake, in which case he is done.) If Alice advances to Stage 2 prematurely (with $\rec = \false$), then the other Alice could not have advanced as well (since her $\rec = \true$). From this point on, Bob sends $\bar{1}$  until the start of the next megablock, and hence this second Alice must remain in Stage 1 the entire time. Then, when the new megablock starts, Bob sends $\bar{0}$ for the rest of the protocol, and this Alice can only either stay in Stage 1 or advance directly to Stage 3, as her $\cnt = 0$. 
    
    In addition, Bob sends $\bar{0}$ for entire blocks only if it is the first time that $\cnt_0 = \cnt_1 = \cnt$ and $x_0[\cnt/2] \not= x_1[\cnt/2]$, or $\cnt_0 = \cnt_1 \pm 1$. In particular, it cannot be the case that $\cnt_0 = \cnt_1 = \cnt = 0\Mod{2}$ and $x_0[\cnt_0/2] = x_1[\cnt_1/2] = 1$, so it cannot be the case that both Alices advance to Stage 2.  
    

    
\item {\bf One of the Alice's is in Stage 3. } 
    In this case, Bob only pays attention to the Alice that is {\em not} in Stage 3 (recall that the Stage 3 Alice ignores him anyway, and continues to send the same bit until the end of the protocol).  His goal is to ensure that the other Alice arrives to Stage 3 with a bit different than that of the current Stage 3 Alice.  This is easy to do since Bob has control over which bit Alice sends in Stage 3 (for example, if the other Alice is in Stage~$1$ he can ensure that she exits Stage~$1$ with $\cnt = 0$ if he wishes her to send $0$, and $\cnt = 1$ if he wishes her to send $1$; a similar strategy works for a Stage~$2$ Alice as well). 

\end{enumerate}

\noindent
We refer the reader to Section~\ref{sec:3/5} for the formal description of the protocol and its analysis. 

\subsection{Discussion}

Both the $\iECC$ constructions presented in this paper involve the same high level idea: progress is made whenever Bob can narrow down Alice's message to two possibilities and Alice can decode Bob's message. Under this template, $\frac35$ is in fact the optimal error resilience: an adversary can stall all progress by erasing $\frac34$ of Alice's messages to confuse Bob between three states, or she can erase all of Bob's messages so that the $\iECC$ reduces to a non-interactive $\ECC$ and then erase $\frac12$ of Alice's messages. Balancing the two attacks gives that an adversary can always succeed in confusing Bob with budget $\frac35$.

We leave open the problem of whether our protocols can be modified to handle more than two worlds, so that progress is made whenever Bob list-decodes Alice's message to $k > 2$ possibilities and Alice uniquely decodes Bob's message. If this were the case, then an adversary could stall progress only if she erases $1 - \frac{1}{2^k}$ of Alice's message. As $k \rightarrow \infty$, the adversary must erase closer and closer to $1$ of Alice's messages. In the limit, the erasure resilience approaches the bound given in Theorem~\ref{thm:intro-2/3}.

\section{Preliminaries and Definitions}
Before we dive into the technical part of our paper, we present important preliminaries on classical error correcting codes, and define an $\iECC$ formally and what it means for one to be resilient to $\alpha$-fraction of errors.


\paragraph{Notation.} In this work, we use the following notations.
\begin{itemize}
    \item The function $\Delta(x, y)$ represents the Hamming distance between $x$ and $y$.
    \item The interval $[0,n]$ for $n\in \bbZ_{\geq 0}$ denotes the integers from $0$ to $n$ inclusive.
    \item The symbol $\perp$ in a message represents the erasure symbol that a party might receive in the erasure model.
    \item When we say Bob $k$-decodes a message, we mean that he list decodes it to exactly $k$ possible messages Alice could have sent in the valid message space.
\end{itemize}

\subsection{Classical Error Correcting Codes}

\begin{definition} [Error Correcting Code]
    An error correcting code ($\ECC$) is a family of maps $\ECC = \{ \ECC_n : \{ 0, 1 \}^n \rightarrow \{ 0, 1 \}^{p(n)} \}_{n \in \bbN}$. An $\ECC$ has \emph{relative distance $\alpha > 0$} if for all $n \in \bbN$ and any $x \not= y \in \{ 0, 1 \}^n$,
    \[
        \Delta \left( \ECC_n(x), \ECC_n(y) \right) \ge \alpha p.
    \]
\end{definition}

where $\Delta$ is the Hamming distance. Binary error correcting codes with relative distance $\approx \frac12$ are well known to exist with linear blowup in communication complexity.

\begin{theorem}[\cite{GuruswamiS00}] \label{thm:ECC-plain}
    For all $\epsilon > 0$, there exists an explicit error correcting code $\ECC_\epsilon = \{ \ECC_{\epsilon, n} : \{ 0, 1 \}^n \rightarrow \{ 0, 1 \}^{p} \}_{n \in \bbN}$ with relative distance $\frac12 - \epsilon$ and with $p = p(n) = O_\epsilon (n)$.
\end{theorem}

A relative distance of $\frac12$ is in fact optimal in the sense that as the number of codewords $N$ approaches $\infty$, the maximal possible relative distance between $N$ codewords approaches $\frac12$. We remark, however, that for small values of $N$, the distance can be much larger: for $N = 2$, the relative distance between codewords can be as large as $1$, e.g. the codewords $0^M$ and $1^M$, and for $N = 4$, the relative distance can be as large as $\frac23$, e.g. the codewords $(000)^M, (110)^M, (101)^M, (011)^M$. As mentioned in Section~\ref{sec:overview}, our constructions leverage this fact that codes with higher relative distance exist for a small constant number of codewords.

We will also need the following important lemma about the number of shared bits between any three codewords in an error correcting code scheme that has distance $\frac12$. 

\begin{lemma} \label{lem:1/4}
    For any error correcting code $\ECC_\epsilon = \{ \ECC_{\epsilon, n} : \{ 0, 1 \}^n \rightarrow \{ 0, 1 \}^{p(n)} \}_{n \in \bbN}$ with relative distance $\frac12 - \epsilon$, and any large enough $n \in \bbN$, any three codewords in $\ECC_{\epsilon, n}$ overlap on at most $\left( \frac14 + \frac32 \epsilon \right) \cdot p$ locations.
\end{lemma}

\begin{proof}
    Consider three codewords $c_1, c_2, c_3$. Suppose that all pairs are relative distance at least $\left( \frac12 - \epsilon \right)$. Let $c_1$ and $c_2$ share $f \le \left( \frac12 + \epsilon \right) \cdot p$ bits, and all three codewords share $e \le f$ bits. Then, note that 
    \begin{align*}
        2p \cdot \left( \frac12 - \epsilon \right)
        &\le \Delta(c_1, c_3) + \Delta(c_3, c_2) \\ 
        &\le 2(f-e) + (p-f) \\
        &= p + f - 2e \\
        &\le \left( \frac32 + \epsilon \right) \cdot p - 2e,
    \end{align*}
    which means that
    \[
        e \le \left( \frac14 + \frac32 \epsilon \right) \cdot p,
    \]
    as claimed.
\end{proof}

Lemma~\ref{lem:1/4} means that assuming that $< \frac34$ of a codeword is erased, the resulting message is list-decodable to a set of size $\le 2$, at least in theory. The following theorem says that this list-decoding is polynomial time. 

\begin{theorem} \cite{Guruswami03,GuruswamiS00} \label{thm:ECC}
    For all $\epsilon > 0$, there exists an explicit error correcting code $\ECC_\epsilon = \{ \ECC_{\epsilon, n} : \{ 0, 1 \}^n \rightarrow \{ 0, 1 \}^{p} \}_{n \in \bbN}$ with relative distance $\frac12 - \epsilon$ and $p = p(n) = O_\epsilon (n)$, and a $\poly_\epsilon(n)$-time decoding algorithm $\DEC_\epsilon = \{ \DEC_{\epsilon,n} : \{ 0, 1 \}^p \rightarrow \mathcal{P}(\{ 0, 1 \}^n) \}_{n \in \bbN}$, such that for any $n \in \bbN$, $x \in \{ 0, 1 \}^n$, and corruption $\sigma$ consisting of fewer than $(\frac34 - \frac32\epsilon) \cdot p$ erasures,
    \[
        \left| \DEC_{\epsilon,n}(\sigma \circ \ECC_{\epsilon, n}(x)) \right| \le 2, \qquad x \in \DEC_{\epsilon,n}(\sigma \circ \ECC_{\epsilon, n}(x)).
    \]
    Furthermore, for all $n \in \bbN$, all codewords $\in \ECC_{\epsilon, n}$ are relative distance $\frac12 - \epsilon$ from the strings $0^p$ and $1^p$, and from the strings $(000)^{p/3}, (011)^{p/3}, (101)^{p/3}, (110)^{p/3}$.\footnote{This last property can be made to hold by taking an appropriate inner code.}
\end{theorem} 

\subsection{Interactive Error Correcting Codes} \label{iecc-def}
We formally define our notion of an \emph{interactive error correcting code ($\iECC$)}. The two types of corruptions we will be interested in are erasures and bit flips. We first start by defining a non-adaptive interactive protocol.

\begin{definition} [Non-Adaptive Interactive Protocol]
    A non-adaptive interactive protocol $\pi = \{ \pi_n \}_{n \in \bbN}$ is an interactive protocol between Alice and Bob, where in each round a single party sends a single bit to the other party. The order of speaking, as well as the number of rounds in the protocol, is fixed beforehand. The number of rounds is denoted $|\pi|$. 
    
\end{definition}

\begin{definition} [Interactive Error Correcting Code]
    An interactive error correcting code ($\iECC$) is a non-adaptive interactive protocol $\pi = \{ \pi_n \}_{n \in \bbN}$, with the following syntax: 
    \begin{itemize}
        \item At the beginning of the protocol, Alice receives as private input some $x \in \{ 0, 1 \}^n$.
        \item At the end of the protocol, Bob outputs some $\hat{x} \in \{ 0, 1 \}^n$.
    \end{itemize}
    We say that $\pi$ is $\alpha$-resilient to adversarial bit flips (resp. erasures) if there exists $n_0 \in \bbN$ such that for all $n > n_0$ and $x \in \{ 0, 1 \}^n$, and for all online adversarial attacks consisting of flipping (resp. erasing) at most $\alpha \cdot |\pi|$ of the total communication, Bob outputs $x$ at the end of the protocol with probability $1$.
    
\end{definition}

\section{Interactive Error Correcting Codes Resilient to $6/11$ Erasures} \label{sec:6/11}

\subsection{Protocol Outline}

Recall that our protocol consists of many equal-length \emph{chunks}, each consisting of $M$ rounds from Alice to Bob followed by $\frac38 M$ rounds from Bob to Alice. We say that all the rounds spoken by a single party within a chunk is a \emph{message}. 

We assume in the following description that Bob never uniquely decodes Alice's message, since then he trivially learns $x$. We also assume that whenever there are exactly two possible values of Alice's message compatible with what Bob received, that they are consistent with previous message pairs received by Bob, 
otherwise Bob can rule out one of the messages and successfully learn~$x$.

\begin{enumerate}
    \item 
        Alice holds a counter $\cnt$ initially set to $0$. Alice begins the protocol by sending $\ECC(x, \cnt)$ to Bob in each chunk.
        
    \item 
        Bob begins the protocol by sending $\bar{0}$ if not otherwise specified. Let $m$ be the (partially erased) message Bob receives from Alice. If Bob cannot list-decode $m$ into at most two options, Bob ignores the message by simply sending the same message as he sent last. The first time that Bob list-decodes $m$ into two possible options, he sets an index $i$ on which they differ, and both possibilities must have $\cnt = 0$.
        
        Whenever he list-decodes into two options, he increments $\cnt$ or tells Alice to start sending him a single bit forever, as follows:
        \begin{itemize}
            \item If the two values of Alice's counter $\cnt_0$ and $\cnt_1$ have both been incremented since the last time he $2$-decoded and are still less than $i$, Bob switches his message to sending $\bar{1}$ if he had previously sent $\bar{0}$, and $\bar{0}$ if he had previously sent $\bar{1}$. 
            \item If the two values of Alice's counter $\cnt_0$ and $\cnt_1$ have both not been incremented since the last time Bob $2$-decoded, Bob continues sending the same message he sent last.
            \item If at any point the two values of Alice's counter have become off-by-$1$, i.e. one has incremented and the other hasn't, Bob sends the codeword $\bar{3}$ for the rest of the protocol to ask Alice for the parity of her counter $\cnt$. Using the answer to this question, Bob can determine whether Alice's true input were $x_0$ or $x_1$.
            \item Otherwise, $\cnt_0 = \cnt_1 = i$, where $i$ is an index for which the two associated values of $x$ differ. Bob then sends $\bar{2}$ for the rest of the protocol to ask Alice for her value of $x[\cnt]$, which will allow him to determine Alice's true input.
        \end{itemize}
    
    \item 
        Whenever Alice unambiguously sees a \emph{change} in Bob's message from a $\bar{0}$ to a $\bar{1}$ or vice versa, she increments her counter by $1$. She does so until she unambiguously receives a $\bar{2}$ or $\bar{3}$, at which point she sends her current value of $x[\cnt]$ or $\cnt~\text{mod}~2$, respectively for the rest of the protocol.
\end{enumerate}

\noindent
The protocol is presented formally in Section~\ref{sec:6/11-protocol}.

\subsection{Formal Protocol} \label{sec:6/11-protocol}

In this section, we describe the $\frac6{11}$ erasure resilient protocol formally. 

\protocol{Interactive Binary One Way Protocol Resilient to $\frac6{11} - \frac{14}{11} \epsilon$ Erasures}{6/11}{
    Let $n$ be the size of the message $x \in \{ 0, 1 \}^n$ that Alice wishes to convey to Bob. Let $\ECC(\cdot, \cdot) : \{ 0, 1 \}^n \times [0,n] \rightarrow \{ 0, 1 \}^M$ be the error correcting code of relative distance $\frac12-\epsilon$ from Theorem~\ref{thm:ECC}, such that every codeword is also distance $\frac12 -\epsilon$ from each of $0^M$ and $1^M$. 
    Let $\bar{0}, \bar{1}, \bar{2}, \bar{3}$ denote length $\frac38 M$ binary strings that have relative distance $\frac23$ from each other (specifically, $(000)^{M/8},(011)^{M/8},(101)^{M/8},(110)^{M/8}$). Our protocol consists of $T = \lceil \frac{n+1}{\epsilon} \rceil$ chunks of Alice sending an $M$ bit message followed by Bob sending a $\frac38 M$ bit message. Throughout the protocol, the parties will choose a response based on which case applies to their received message $m$; if multiple apply, they choose the first case on the list.
    
    \begin{center}
    {\bf \fbox{Alice}}
    \end{center}
    
        In addition to $x$, Alice has an internal state consisting of
        \begin{itemize}
            \item A counter $\cnt$ that at the beginning of the protocol is set to $0$.
            \item An internal state $\mes$, originally set to $\bar{0}$, representing Bob's most recent message that successfully got through to her.
        \end{itemize}
        
        She begins the protocol by sending $\ECC(x, \cnt)$ as the first message. Before every future message, she takes note of the latest message ($\frac38 M$ rounds) from Bob as $m \in \{0,1,\bot\}^{3M/8}$, and determines her response as follows. Note that if $<\frac23$ of the symbols in $m$ are erasures, there is only one value of $s\in \bar{0}, \bar{1}, \bar{2}, \bar{3}$ consistent with the message $m$.
        
        \begin{caseof}
            \case {$\geq \frac23$ of the symbols in $m$ are $\bot$.}
            {Alice sends the same message as in the previous chunk.}
            
            \case{$m$ uniquely decodes to $s\in \{\bar{0}, \bar{1}\}$.}{
                If $s \not= \mes$, Alice increments $\cnt$ by $1$ and sets $\mes$ to $s$. She then sends $\ECC(x, \cnt)$. (In other words, $\cnt$ is the number of times that Alice detected a flip in Bob's messages from $\bar{0}$ to $\bar{1}$ or vice versa.)
                
                If $s = \mes$, Alice sends the same message as in the previous chunk. 
            }
            
            \case{$m$ uniquely decodes to $s=\bar{2}$.}{
                Alice sends $1^M$ if $x[\cnt] = 1$ and $0^M$ if $x[\cnt] = 0$ in all subsequent chunks, ignoring any future instructions.
            }
            
            \case{$m$ uniquely decodes to $s=\bar{3}$.}{
                Alice sends $(\cnt ~\text{mod}~ 2)^M$ in all subsequent chunks, once again ignoring any future instructions.
            }
        
        \end{caseof}
    
    \begin{center}
    {\bf \fbox{Bob}}
    \end{center}
    
        Bob holds a variable $\hat{x}$, initially set to $\emptyset$, that will be updated with his final output either at the end of the protocol or once he has unambiguously learns Alice's value of $x$. Once $\hat{x}$ is set to a value ($\neq \emptyset$), it will not be updated again. That is, Bob ignores any future instructions to update it. At the end of the protocol, Bob outputs $\hat{x}$. If at any point in the protocol $\hat{x}$ has already been set, Bob may send Alice any arbitrary message, say $\bar{1}$. 
        
        Bob also keeps track of the following values:
        \begin{itemize}
            \item Two values $\hat{x}_0$ and $\hat{x}_1$, to be set when Bob $2$-decodes Alice's message for the first time.
            \item A fixed index $i$ where $\hat{x}_0$ and $\hat{x}_1$ differ, set as soon as $\hat{x}_0$ and $\hat{x}_1$ are known.
            \item $\mes$, representing the last message ($\bar{0}$ or $\bar{1}$) he sent Alice; $\mes$ is originally set to $\bar{0}$.
            \item $\last$, representing the last value Bob heard of Alice's counter, originally set to $0$.
            \item A value $\ques\in\{2,3\}$, indicating whether Bob wants to know the value of $x[i]$ or the parity of Alice's counter, respectively. This is only set when Bob transitions from Phase 1 to Phase 2.
            \item A bit $\parity \in \{ 0, 1 \}$, used in Phase 3 only if $\ques$ is set to $3$, and is set at the same time as $\ques$. 
        \end{itemize}
    
        Each chunk, Bob's outgoing message is one of four codewords: $\bar{0}$, $\bar{1}$, $\bar{2}$, or $\bar{3}$. He begins the protocol in Phase 1 and at some point may transition to Phase 2, but can never go back to Phase 1. In Phase 1, his goal is to increment Alice's counter value and may only ever send $\bar{0}$ or $\bar{1}$. In Phase 2, his goal is to tell Alice to switch to sending only a single final bit, and in this phase he may only ever send either $\bar{2}$ or $\bar{3}$, but not both.
    
        \begin{enumerate}[align=left, leftmargin=*, label={\bf Phase \arabic*:}]
        \item 
            In Phase 1, Bob's message to Alice is always either $\bar{0}$ or $\bar{1}$. Let Alice's most recent message be $m \in \{ 0, 1, \bot \}^M$. Bob determines his next message depending on which of the following cases $m$ falls under. Note that if $<\frac34 - \frac32 \epsilon$ of the symbols in $m$ have been erased, there are at most two values of $s\in \{\ECC(x,\cnt)\}_{x, \cnt}$ consistent with the message $m$.
            
            \begin{caseof}
            
            \case {$\geq \frac34-\frac32\epsilon$ of the symbols in $m$ are $\bot$.}{
                Bob simply sends Alice $\mes$ again.
            }
            
            \case {$m$ is $\leq 2$-decoded where at most one element is of the form $\ECC(x, \cnt)$. 
                There cannot be zero values of this form, since Alice only sends messages of the form $\ECC(x,\cnt)$ while she has heard only $\bar{0}$'s and $\bar{1}$'s. Bob can therefore uniquely decode $m$ to $\ECC(x, \cnt)$, which must be Alice's true message. He then sets $\hat{x} \gets x$.
            }
            
            \case {$m$ is $2$-decoded to two states $\{ \ECC(x_0, \cnt_0), \ECC(x_1, \cnt_1) \}$ for the first time.}{
                In this case, the only value Bob has sent so far is $\bar{0}$. Therefore, if $\cnt_0 \not= 0$ or $\cnt_1 \not= 0$ (in which case exactly one is true as one must be Alice's real state), then Bob sets $\hat{x} \gets x_b$ where $b$ is such that $\cnt_b = 0$. 
                
                Otherwise, $\cnt_0 = \cnt_1 = 0$. For $b \in \{ 0, 1 \}$, Bob sets $\hat{x}_b \gets x_b$. He also sets $i$ to be an index where $\hat{x}_0$ and $\hat{x}_1$ differ. He sets $\mes = \bar{1}$ and sends $\mes$.
            }
            
            \case {$m$ is $2$-decoded to two states $\{ \ECC(x_0, \cnt_0), \ECC(x_1, \cnt_1) \}$, not for the first time.}{
                Bob switches the order of the two states if need be, so that at least one of $x_0 = \hat{x}_0$ or $x_1 = \hat{x}_1$.
                \begin{itemize}[leftmargin=*]
                    \item If for some $b \in \{ 0, 1 \}$ it holds that $x_b \not= \hat{x}_b$ or $\cnt_b \not\in \{ \last, \last + 1 \}$, then Bob sets $\hat{x} \gets x_{1-b}$. 
                    \item If $\cnt_0 = \cnt_1 = \last$, he sends $\mes$.
                    \item If $\cnt_0 = \cnt_1 = \last + 1$, he sets $\last := \cnt_0 = \cnt_1$. 
                    If the new value $\last$ is equal to the index $i$, Bob sets $\ques=2$ and moves to Phase 2. 
                    Otherwise, $\last < i$. Bob sets $\mes$ to be the logical not of itself, that is, if $\mes = \bar{0}$ the new value of $\mes$ is $\bar{1}$, and vice versa. He sends $\mes$.
                    \item If $\cnt_0 \not= \cnt_1$, then $|\cnt_0 - \cnt_1| = 1$. Bob sets $\ques=3$ and $\parity = (\cnt_1~\text{mod}~2)$, and moves to Phase 2.
                \end{itemize}
            }
            \end{caseof}
            
            If at the end of the protocol Bob is still in Phase 1, he sets $\hat{x}$ to a random value of $x$.
            
        \item 
            Bob always enters Phase 2 with a value of $\ques\in\{2,3\}$ permanently set. He sends $\overline{\ques}$ each chunk for the rest of the protocol. 
        
            At the end of the protocol, let $d$ denote Bob's most recently received bit from Alice.
            If $\ques = 2$, he sets $\hat{x} \gets \hat{x}_b$, where $\hat{x}_b$ is such that $\hat{x}_b[i] = d$.
            If $\ques = 3$, he sets $\hat{x} \gets \hat{x}_b$, where $b$ is $1$ if $d = \parity$ and $0$ otherwise. 
    \end{enumerate}
}

\subsection{Analysis}

\begin{theorem}
    Protocol~\ref{prot:6/11} is resilient to a $\frac6{11}-\frac{14}{11}\epsilon$ fraction of erasures. For an input of size $n$, the total communication is $O_\epsilon(n^2)$. Alice and Bob run in $\poly_\epsilon(n)$ time.
\end{theorem}

\begin{proof}
    The number of communicated bits is $T \cdot \frac{11}{8} M = \lceil \frac{n+1}\epsilon \rceil \cdot O_\epsilon(n) = O_\epsilon(n^2)$. The runtime of Alice and Bob is governed by the time it takes to list-decode, which by Theorem~\ref{thm:ECC} is $\poly_\epsilon(n)$. We focus on proving error resilience.
    
    Suppose that Bob outputs an incorrect value of $x$. We will show that the adversary must have erased at least $\frac6{11} - \frac{14}{11}\epsilon$ of the communication.
    
    First, we claim that if Alice ever uniquely decodes Bob's message in a chunk $R$ to be $\bar{2}$ or $\bar{3}$ and Bob hears at least one bit from Alice in a chunk after $R$, then Bob will output the correct value of Alice's input $x$. This is because the final bit will convey to him either the value of $x[i]$ or the parity of $\cnt$, which allows him to distinguish between $\hat{x}_0$ and $\hat{x}_1$. Furthermore, any bit Alice sends after the $R$'th chunk must in fact be $x[i]$ if $\ques = 2$, or $\cnt~\text{mod}~2$ if $\ques = 3$, since she had previously uniquely decoded Bob's message to be $\bar{2}$ or $\bar{3}$, so Bob must have actually sent $\bar{2}$ or $\bar{3}$ respectively.
    
    Let $R$ be the first chunk in which Alice uniquely decodes Bob's message to be $\bar{2}$ or $\bar{3}$, and if such a chunk does not exist then let $R=T$. The argument above implies that if Bob outputs the incorrect value of $x$, it must be the case that either none of Alice's messages after chunk $R$ got through to Bob.
    
    By the definition of our protocol, in the first $R$ chunks, Alice sends $\ECC(x, \cnt)$ for some $\cnt\in\{0,1\,\ldots,i\}$. Since we assumed that Bob outputs $\hat{x} \not= x$, it must be the case that none of these messages can be uniquely decoded, and in particular at least $\frac12 - \epsilon$ of each of Alice's messages must be erased, otherwise Bob uniquely decodes Alice's message and sets $\hat{x}$ correctly. Let $S$ be the number of these $R$ chunks in which at least $\frac34 - \frac32 \epsilon$ of Alice's messages are corrupted. In the other $R - S$ chunks, between $\frac12-\epsilon$ and $\frac34-\frac32\epsilon$ of Alice's messages are corrupted. We next argue that in at most $i+1$ of these $R-S$ chunks, Bob's messages to Alice have a unique decoding.  This is the case since whenever Alice uniquely decodes Bob's message, she increments her counter (or has just received a $\bar{2}$ or $\bar{3}$ and knows to begin sending a single bit for the rest of the protocol), and this change is heard by Bob since he can $\le 2$-decode Alice's message in each of these $R-S$ chunks.  As we have established, Alice's counter never exceeds $i$, which implies that in at most $i+1 \le n+1$ of these $R-S$ chunks, Bob's messages to Alice have a unique decoding. In the other $\ge R - S - n - 1$ chunks, Bob's message to Alice is at least $\frac23$ corrupted. This gives a total corruption rate of at least
    \begin{align*}
        &\frac{(T - R) \cdot M + S \cdot (\frac34-\frac32\epsilon) M + (R - S) \cdot (\frac12-\epsilon) \cdot M + (R - S - n - 1) \cdot \frac23 \cdot \frac38 M}{T \cdot \frac{11}{8} M} \\
        =& \frac{T - \frac\epsilon2 S - (\frac14 + \epsilon) R - \frac14 (n+1)}{\frac{11}8 T} \\
        \ge& \frac{(\frac34 - \frac32\epsilon) T  - \frac14 (n+1)}{\frac{11}8 T} \\
        \ge& \frac{6}{11} - \frac{14}{11} \epsilon,
    \end{align*}
    where in the last step we use that $T = \lceil \frac{n+1}{\epsilon} \rceil$.
\end{proof}

\section{Interactive Error Correcting Codes Resilient to $3/5$ Erasures} \label{sec:3/5}

We recommend the reader understand the protocol achieving a $\frac6{11}$-resilience to erasures given in Section~\ref{sec:6/11} before reading this section. 

\subsection{Protocol Overview}

We refer the reader to the technical overview for a more comprehensive introduction to our protocol. In this section, we recall a couple changes made for our new protocol as compared to the $\frac6{11}$ protocol, then give an outline of the protocol.

First, we introduce two new layers of round grouping on top of the chunks (which we recall are two messages, one from Alice and one from Bob). The first is the \emph{block}, which consists of $O_\epsilon(n)$ chunks. The second is the \emph{megablock}, which consists of $O_\epsilon(1)$ blocks. Alice and Bob increment throughout a megablock at most once per block, as opposed to once per chunk as in the $\frac6{11}$. At the end of a megablock, Alice resets her counter. 

Furthermore, Bob now attempts to increment $\cnt$ to the value $2i$ instead of $i$. Alice now understands the \emph{value} question to be asking for the value of $x[\cnt/2]$, and the \emph{parity} question to be asking for the parity of $\cnt$.

Alice can be in one of three stages: Stage 1, in which she increments $\cnt$; Stage 2, in which she increments a new counter $\knt$ to learn Bob's question; and Stage 3, in which she simply sends the same bit $\beta$ for the rest of the protocol. Alice begins the protocol in Stage 1. At some point, she either advances directly to Stage 3 or first advances to Stage 2 before advancing further to Stage 3.

Meanwhile, Bob can be in one of three phases depending on the stages of the two Alices he decodes to. He begins in Phase 1, in which both Alices are in Stage 1. If at some point he decodes a Stage 1 Alice and a Stage 2 Alice, he transitions to Phase 2. Otherwise, if at some point he decodes a Stage 3 Alice, he transitions to Phase 3. Once he has entered either Phase 2 or 3, he stays there for the rest of the protocol. 


\paragraph{The protocol.} The final protocol is as follows. We assume that whenever Bob list-decodes Alice's message to two states that both states are consistent with previous pairs of states, that is, they are possible values of Alice's current state if she had been in the previous state and had seen some subset of Bob's message since the last $2$-decoding. Otherwise, if Bob list-decodes to states such that only one is consistent with a past state, he learns $x$ and doesn't need to go through with the rest of the protocol.

\begin{enumerate}
    
    \item At the start of each megablock, Alice and Bob engage in an incrementation procedure in which Bob attempts to increment Alice's counter $\cnt$ to the value $2i$. At the start of each megablock, Alice resets $\cnt$, and Bob, knowing this, restarts the incrementation. 
    
    \item Eventually, one of two things will happen.
    
    \begin{itemize}
        \item Bob begins sending $\bar{0}$'s for the rest of the megablock to tell both Alices to advance from Stage 1 to Stage 2 or 3.

        \item At some premature point, before the conditions for asking both Alices to advance have been reached, Bob sees that exactly one of the Alices has already advanced. This is possible if at some point Bob sees that one Alice has received a $\bar{1}$ while the other has not, since in that case he sends $\bar{0}$ for the rest of the block.
    \end{itemize}
    
    \item Alice advances to Stage 2 or possibly skips directly to Stage 3 if the first message she receives in a block is a $\bar{0}$. In Stage 2, Alice attempts to figure out whether Bob wants to know the answer to the \emph{parity} or \emph{value} question. In Stage 3, she sends a single bit $\beta$ representing the answer to this question forever.
    
    In order to figure out which of the stages to advance to, she looks at $x[\cnt/2]$ and $\cnt~\text{mod}~2$. If they are both the same bit $b$, she can automatically proceed to Stage 3, with $\beta=b$. If $\cnt$ is odd, she knows Bob is asking the \emph{parity} question, so she can still move to Stage 3. Only if $x[\cnt/2]=1$ and $\cnt~\text{mod}~2=0$ does she enter Stage 2. 
    
    In Stage 2, starting with the next megablock after first advancing to Stage 2, Alice increments $\knt$ instead of $\cnt$, resets $\knt$ (and not $\cnt$) at the start of every megablock, and advances to Stage 3 when she hears $\bar{0}$ as the first unerased message of a block. At that point, if $\knt=0$ she advances to Stage 3 with $\beta=x[\cnt/2]=1$, and otherwise with $\beta=\cnt~\text{mod}~2=0$.
    
    \item When Bob sees that at least one Alice has advanced to Stage 2 or 3, he sends $\bar{1}$'s for the rest of the megablock, and changes his strategy in the next megablock. Note that both versions of Alice cannot be in Stage 2, since then they would both have $\cnt ~\text{mod}~ 2 =0$ and $x[\cnt/2] = 1$, but Bob only allows both Alices to advance if the two Alices have different values of $\cnt~\text{mod}~2$ or $x[\cnt/2]$. 
    
    \begin{caseof}
        \case {One Alice is in Stage 1, and the other is in Stage 2.} 
            {It is the beginning of the megablock so the Alice in Stage 1 has $\cnt = 0$ and the Alice in Stage 2 has $\knt = 0$. Bob simply sends $\bar{0}$ for the rest of the protocol, so that when the Alices advance to Stage 3, they do so with opposite bits: the Alice in Stage $1$ answers $0$, and the Alice in Stage $2$ answers $x[\cnt/2] = 1$. This is known as Phase 2.}
        \case {One Alice is in Stage 1 or 2, and the other Alice is in Stage 3.} 
            {The answer $\beta \in \{ 0, 1 \}$ of the Stage 3 Alice is fixed, so Bob can focus on incrementing the remaining Alice's counter to a value for which her answer to Bob's question is $1-\beta$. This is known as Phase 3.}
    \end{caseof}
    
    \item At the end of the protocol, Bob should have coordinated both versions of Alice to send opposite bits forever; hearing any one bit from Alice now suffices to deduce $x$.
\end{enumerate}

\noindent We provide the formal protocol in the next section.



\subsection{Formal Protocol} \label{sec:3/5-formal}

\protocol{Interactive Binary One Way Protocol Resilient to $\frac35 - \epsilon$ Erasures}{3/5}{
    \setlength{\parindent}{1em}

    Let $n$ be the size of the message $x \in \{ 0, 1 \}^n$ that Alice wishes to convey to Bob. Let 
    \begin{align*}
        \ECC(x, \cnt, \sent, \rec, \knt, \stg) : \ & \{ 0, 1 \}^n \times [0,n] \times \{\true, \false \} \times \{ \true, \false \} \times \{-1,0,1\} \times \{ \true, \false \} \\ 
        & \rightarrow \{ 0, 1 \}^{4M}
    \end{align*}
    be an error correcting code of relative distance $\frac12-\epsilon$ such that each codeword is also relative distance $\ge \frac12 - \epsilon$ from each of $0^{4M}, 1^{4M}$, as given in Theorem~\ref{thm:ECC}. Also, let $\bar{0}=0^M$ and $\bar{1}=1^M$. Our protocol consists of $A =\lceil{ \frac{1}{\epsilon}\rceil}$ megablocks of $B = \lceil{\frac{n}{\epsilon}\rceil}$ blocks, each consisting of $C =\lceil{ \frac{1}{\epsilon}\rceil}$ chunks of $5M$ rounds, made of Alice sending a $4M$ bit message followed by Bob sending a $M$ bit message. Throughout the protocol, the parties will choose a response based on which case applies to their received message $m$; if multiple apply, they choose the first case on the list.
    
    \begin{center}
    {\bf \fbox{Alice}}
    \end{center}
        
    Alice's message will always be from the following set: $$\{ \ECC(x,\cnt,\sent,\rec,\knt,\stg) \}_{x,\cnt,\sent,\rec,\knt,\stg} \cup \{ 0^{4M}, 1^{4M} \}$$ which has relative distance $\ge \frac12 - \epsilon$. She tracks certain values explicitly, namely the \emph{stage} of the protocol she is in, and the following variables:
    \begin{itemize}
    \item 
        A counter $\cnt$ taking values in $[0, n]$, initially set to $0$.
    \item 
        A bit $\sent \in \{\true,\false\}$ denoting whether she has received a $0$ from Bob after sending this particular value of $\cnt$ and later $\knt$ (i.e. whether she has received \emph{confirmation} that $\cnt$/$\knt$ was successfully sent to Bob) within this block. Initially $\sent$ is set to $\true$.
    \item 
        A bit $\rec \in \{\true,\false\}$ denoting whether she has \emph{received} a $1$ so far this block. $\rec$ is initially set to $\false$.
    \item 
        A second counter $\knt$ taking values in $\{ -1, 0, 1 \}$. It will be used in Stage 2 to determine which question Bob wishes answered. It is initially set to $-1$ when Alice is in Stage 1. When Alice advances to Stage 2, she sets $\knt \gets 0$. 
    \item 
        A boolean $\stg$ indicating whether she advanced to stage 2 within this megablock. It is initially set $\false$. $\stg$ is used to ensure Alice does not advance stages twice within the same megablock.
    \end{itemize}
        
        Alice acts differently according to which of three stages she is in. In Stage 1, she increments her counter $\cnt$; in Stage 2, she increments $\knt$ to learn whether Bob wants her to answer the parity or value question; and in Stage 3 she sends only the single bit $\beta$ equal to the answer to Bob's question for the rest of the protocol. Alice may go directly from Stage 1 to 3, or pass through Stage 2 in the middle.
        
    \begin{enumerate}[align=left, leftmargin=*, label={\bf Stage \arabic*:}]
    
    \item 
        At the beginning of each megablock, she resets all the above variables to their initial states.
            
        At the beginning of each block, she resets $\rec = \false$. As her first message of each block, she sends $\ECC(x, \cnt, \sent, \rec, \knt = -1, \stg=\false)$. Let $m \in \{ 0, 1, \perp \}^M$ be the most recent message from Bob; for each remaining message in the block, she determines what to send Bob based on which case $m$ falls under. Note that if $m$ is not entirely erased, then Alice can uniquely decode $m$ to one of $\bar{0}$ and $\bar{1}$.
        
        \begin{caseof}
            
            \case {All of the symbols in $m$ are $\perp$.}{
                Alice sends the same message as the last chunk: $\ECC(x, \cnt, \sent, \rec, \knt = -1, \stg = \false)$.
            }
                
            \case {$m$ decodes uniquely to $\bar{1}$.}{
                Alice sets $\rec \gets \true$. If $\sent = \true$, she increments $\cnt$ and sets $\sent \gets \false$. She sends her updated value of $\ECC(x,\cnt,\sent,\rec,\knt = -1, \stg=\false)$.
            }
                
            \case {$m$ decodes uniquely to $\bar{0}$, and $\rec = \true$ (i.e. she's already seen a $1$ this block).}{
                Alice sets $\sent \leftarrow \true$. She sends $\ECC(x,\cnt,\sent,\rec,\knt = -1, \stg=\false)$.
            }
                
            \case {$m$ uniquely decodes to $\bar{0}$, and $\rec = \false$ (i.e. she has seen no messages this block and the first message received is a $\bar{0}$).}{
                If $\cnt$ is odd, she advances to Stage 3 of the protocol with $\beta = 1$. 
                
                Otherwise, $\cnt$ is even, and if $\cnt = 0$ or $x[\cnt/2] = 0$, she advances to Stage 3 with $\beta=0$. 
                
                Else, $\cnt$ is even and $x[\cnt/2] = 1$. She advances to Stage 2. 
            }
            
        \end{caseof}
            
    \item 
        Note that $x[\cnt/2]=1$ and $\cnt~\text{mod}~2=0$ in order for Alice to enter this stage.
        
        As soon as Alice enters Stage 2, she sets $\knt \gets 0$ and $\stg=\true$. For the rest of the megablock, she sends $\ECC(x, \cnt, \sent, \rec, \knt = 0, \stg = \true)$, ignoring Bob's messages and not updating any of her values.
        
        At the start of each successive megablock, Alice resets $\sent \gets \true$, $\rec \gets \false$, $\knt \gets 0$, and $\stg \gets \false$. We point out that she never alters $\cnt$ again. At the beginning of each block, she resets $\rec \gets \false$. She begins the block by sending $\ECC(x, \cnt, \sent, \rec, \knt, \stg)$. Let $m \in \{ 0, 1, \perp \}^M$ be the most recent message from Bob; she determines what to send next based on the following cases. As before, if at least one symbol in $m$ is not $\perp$, then Bob's original message can only be one of $\bar{0}, \bar{1}$.
        
        \begin{caseof}
            \case {All of the symbols in $m$ are $\perp$.}{
                Alice sends $\ECC(x, \cnt, \sent, \rec, \knt, \stg = \false)$.
            }
                
            \case {$m$ uniquely decodes to $\bar{1}$}{  
                Alice sets $\rec \gets \true$. If $\sent = \true$, she increments $\knt$ and sets $\sent \gets \false$. 
                She sends her updated value of $\ECC(x, \cnt, \sent, \rec, \knt, \stg = \false)$.
            }
                
            \case {$m$ uniquely decodes to $\bar{0}$, and $\rec = \true$ (i.e. she's already seen a $1$ this block).}{
                Alice sets $\sent \leftarrow \true$. She sends $\ECC(x, \cnt, \sent, \rec, \knt, \stg = \false)$.
            }
                
            \case {$m$ uniquely decodes to $\bar{0}$, and $\rec = \false$ (i.e. she has seen no messages this block and the first message received is a $\bar{0}$).}{
                If $\knt = 0$, she advances to Stage 3 of the protocol with $\beta = x[\cnt/2] = 1$. 
                
                Otherwise, if $\knt = 1$, she advances to Stage 3 of the protocol with $\beta = (\cnt~\text{mod}~2) = 0$.
            }
        \end{caseof} 
            
    \item 
        When Alice reaches this stage, she has noted a bit $\beta$. She sends this for the rest of the protocol; all her messages from here on out are irreversibly $\beta^{4M}$. 
        
    \end{enumerate}
        
    \begin{center}
    {\bf \fbox{Bob}}
    \end{center}
    
        Bob holds a variable $\hat{x}$, initially set to $\emptyset$, representing his final output. It will either be set at the end of the protocol or sometime during the protocol if he unambiguously learns Alice's value of $x$. In either case, once set it is final, and at the end of the protocol, Bob outputs $\hat{x}$. We remark that once $\hat{x}$ is set, Bob may behave arbitrarily for the rest of the protocol: let us say that he simply sends $\bar{1}$ for each message thereafter and does no further updates to any of his internal variables.
        
        Bob tracks which phase he is in explicitly, and he holds certain values to help him perform consistency checks of 2-decoded Alice's messages with previous pairs:
        \begin{itemize}
            \item $\hat{x}_0$ and $\hat{x}_1$ storing two possible values of $x$ Alice holds.
            \item $S_0$ and $S_1$ storing all possible messages Alice could send next that are consistent with previous 2-decoded pairs of messages, assuming she has received some subset of messages from Bob since his last 2-decoding, corresponding to her holding the input $\hat{x}_0$ and $\hat{x}_1$ respectively.
        \end{itemize}
        
        The values $\hat{x}_0, \hat{x}_1, S_0, S_1$ are initialized and updated as follows. He initializes $\hat{x}_b, S_b$ immediately upon receiving a $2$-decodable message from Alice. If he has not yet set $\hat{x}$, he updates $S_b$ each time he 2-decodes Alice's message and also after each each message he sends to Alice. 
        
        \begin{itemize}[align=left]
        
        \item[\hspace{1.5em}\textbf{Initializing $\hat{x}_b, S_b$:}] 
            The first time in the entire protocol that Bob receives a message from Alice that has $< \frac34 - \frac32\epsilon$ erasures, Bob has only sent $\bar{1}$ so far, and he sets $\{\hat{x}_b, S_b\}_{b\in\{0,1\}}$, as follows: 
            \begin{itemize}
            \item 
                If Bob $1$-decodes Alice's message to $\ECC(x, \cnt, \sent = \false, \rec, \knt=-1,\stg=\false)$, where $(\cnt, \rec) \not= (0, \true)$, or both messages are of the form $\ECC(x_b, \cnt_b, \sent_b = \false, \rec_b, \knt_b=-1,\stg_b=\false)$ and $x_0 = x_1 = x$, Bob sets $\hat{x} \gets x$. 
            \item 
                Otherwise, if Bob $2$-decodes Alice's message to
                \begin{align*}
                    \left\{ \begin{aligned} \ECC(x_0, \cnt_0, \sent_0 = \false, \rec_0, \knt_0 = -1, \stg_0 = \false), \\
                    \ECC(x_1, \cnt_1, \sent_1 = \false, \rec_1, \knt_1 = -1, \stg_1 = \false) 
                    \end{aligned} \right\}
                \end{align*} 
                both satisfying $(\cnt_b, \rec_b) \not= (0, \true)$ such that $x_0 \not= x_1$, Bob sets $\hat{x}_0 \gets x_0$ and $\hat{x}_1 \gets x_1$. He also sets $S_0 \gets \{ \ECC(x_0, \cnt_0, \sent_0, \rec_0, \knt_0, \stg_0) \}$ and $S_1 \gets \{ \ECC(x_1, \cnt_1, \sent_1, \rec_1, \knt_1, \stg_1)\}$. 
            \end{itemize}
            
            
        \item[\hspace{1.5em}\textbf{Updating $S_b$:}] 
            Bob updates the sets $S_0, S_1$ whenever he receives a message from Alice that is $< \frac34-\frac32\epsilon$ erased, and also \emph{after} each message he sends. 
            \begin{itemize}
            \item 
                Whenever Bob receives a message $m$ from Alice with fewer than $\frac34 - \frac32\epsilon$ erasures, if he does not set $\hat{x}$ in response, he must have $2$-decoded $m$ to two messages $m_0 \in S_0, m_1 \in S_1$. He sets $S_0 = \{ m_0 \}$ and $S_1 = \{ m_1 \}$.
                
            \item 
                After sending a message $m_B$, Bob updates $S_b$ as follows. For each $m_A \in S_b$, he computes the $\le 2$ next possible messages that Alice would send if she had last sent $m_A$ and now either hears or doesn't hear $m_B$. The new set $S_b$ consists of all such possible new messages. 
                
                We remark that the message $m_A$ is sufficient to compute the possible new messages: if $m_A = \ECC(\hat{x}_b, \cnt_b, \sent_b, \rec_b, \knt_b, \stg_b)$, then the variables $\hat{x}_b, \cnt_b, \sent_b, \rec_b, \knt_b, \stg_b$ are sufficiently to simulate Alice's behavior whether she hears $m_B$ or not; and if $m_A = \beta^{4M}$ for some $\beta \in \{ 0, 1 \}$, then the only possible new message that Alice sends is the same message $m_A = \beta^{4M}$. 
            \end{itemize}
        \end{itemize}
        
        
        Each of Bob's messages throughout the protocol is either $\bar{0}$ or $\bar{1}$. Bob begins the protocol in Phase 1, and eventually transitions to either Phase 2 or Phase 3, where he stays for the rest of the protocol. Bob behaves differently depending on which phase he's in. 

        \begin{enumerate}[align=left, leftmargin=*, label={\bf Phase \arabic*:}]
        
        \item 
            In Phase 1, Bob's goal is to increment Alice's counter so that it reaches $2i$, where $i$ is an index for which $x_0[i] \not= x_1[i]$. If this goal is reached, i.e. $\cnt_0 = \cnt_1 = \cnt$ and $x_0[\cnt/2] \not= x_1[\cnt/2]$, or the counters become misaligned, i.e. $\cnt_0 = \cnt_1 \pm 1$, he switches from incrementing Alice's counter to sending her $\bar{0}$'s for the rest of the megablock. If he ever sees that one of the two Alices has advanced to Stage 2 or 3, he transitions to Phase 2 or Phase 3 depending on the stages of the two Alices he sees. 
        
            At the start of every megablock in which Bob is still in Phase 1 at the beginning, Bob sends $\bar{1}$. He determines the message he sends next based on the following cases for the message $m \in \{ 0, 1, \perp \}^{4M}$ he most recently received from Alice:
            \begin{caseof}
            \case {$m$ has at least $\frac34 - \frac32\epsilon$ $\perp$'s.}{
                Bob repeats his last message, unless it is the first message of a block, in which case he sends $\bar{1}$.
            }
            \case {Bob $\leq 2$-decodes $m$ to $\{ m_0, m_1 \}$, and $S_b \cap \{ m_0, m_1 \} = \emptyset$ for some $b \in \{ 0, 1 \}$.}{
                Bob sets $\hat{x}\gets\hat{x}_{1-b}$.
            }
            \end{caseof}
        
            In all remaining cases, Bob $2$-decodes $m$ to two messages $m_0 \in S_0$ and $m_1 \in S_1$. 
            
            \begin{caseof}
            \setcounter{casenum}{3}
            
            \case {One of the worlds has already advanced, that is, for some $b \in \{ 0, 1 \}$, either $m_b = \ECC(\hat{x}_b, \cnt_b, \sent_b, \rec_b, \knt_b, \stg_b)$ with $\knt_b \in \{ 0, 1 \}$, or $m_b \in \{ 0^{4M}, 1^{4M} \}$.}{
                Bob sends $\bar{1}$ for every message for the rest of the megablock, ignoring any instructions to send any other messages. He still updates $S_0,S_1$ and checks consistency of $m$ with $S_b$ (Case 2). 
            
                If either $m_0$ or $m_1$ is in $\{ 0^{4M}, 1^{4M} \}$, Bob transitions to Phase 3 at the beginning of the next megablock. Otherwise, if $m_b = \ECC(\hat{x}_b, \cnt_b, \sent_b, \rec_b, \knt_b \in \{ 0, 1 \}, \stg_b)$ for some $b \in \{ 0, 1 \}$, he transitions to Phase 2 at the beginning of the next megablock.
            }
            \end{caseof} 
            
            For the rest of the cases, $m_0 = \ECC(\hat{x}_0, \cnt_0, \sent_0, \rec_0, \knt_0 = -1, \stg_0 = \false)$ and $m_1 = \ECC(\hat{x}_1, \cnt_1, \sent_1, \rec_1, \knt_1 = -1, \stg_1 = \false)$. 
            
            \begin{caseof}
            \setcounter{casenum}{4}
            
            \case {$\cnt_0 = \cnt_1 = 2i$ or $\cnt_0 \not= \cnt_1$.}{
                Bob sends only $\bar{0}$'s for the rest of the megablock, ignoring any instructions to send other messages. He still updates $S_0,S_1$, checks compatibility of $S_b$ with $m$ (Case 2), and transitions to Phase 2 or 3 if appropriate (Case 3).
            }
            
                
            
                
                
            \case {$\cnt_0 = \cnt_1 \not= 2i$ and $\rec_0 = \rec_1 = \false$.}{
                Bob sends $\bar{1}$.
            }
                
            \case {$\cnt_0 = \cnt_1 \not= 2i$ and $\rec_0 \vee \rec_1 = \true$.}{
                Bob sends $\bar{0}$.
            }
                
                
            \end{caseof}

            At the end of the megablock, Bob begins the next megablock in Phase 1 again, unless he was specified to move to Phase 2 or 3. 
            
            
        \item 
            Upon entering Phase 2 at the beginning of a megablock, Bob has previously decoded to two possible Alices in the previous megablock, at least one of which was in Stage 2 ($\knt \in \{ 0, 1 \}$). In fact, by Lemma~\ref{lemma:3/5-S0S1},
            this is true for \emph{exactly one} of the two Alices. That is, one Alice was in Stage 1 ($\knt = -1$) and the other was in Stage 2 $(\knt \in \{ 0, 1 \}$). W.l.o.g. suppose that the Stage 2 Alice corresponded to $\hat{x}_{1}$. Since Bob sent $\bar{1}$ in every message after this 2-decoding for the rest of the megablock, the real Alice must still be in the same stage as she was in when Bob 2-decoded. Now, Bob simply sends $\bar{0}$ in every message for the remainder of the protocol. 
            
            He continues list-decoding Alice's messages with $< \frac34-\frac32\epsilon$ erasures to $\{ m_0, m_1 \}$. If $S_b \cap \{ m_0, m_1 \} = \emptyset$ for some $b \in \{ 0, 1 \}$, he sets $\hat{x} \gets \hat{x}_{1-b}$.
        
            At the end of the protocol, let $b$ be the bit Bob last received. He sets $\hat{x} \gets \hat{x}_{b}$.
            
        \item 
            When Bob enters Phase 3, it is at the beginning of a megablock. He has previously list-decoded to two possible messages, at least one of which was of the form $\beta^{4M}$ for some $\beta \in \{ 0, 1 \}$. Note that in fact exactly one of the messages must have been of the form $\beta^{4M}$ since $0^{4M}$ and $1^{4M}$ have no overlapping bits. W.l.o.g. suppose that when Bob list-decoded to the two messages, the two messages were $m_0 = \ECC(\hat{x}_0, \cnt_0, \sent_0, \rec_0, \knt_0, \stg_0) \in S_0$ and $m_1 = \beta_1^{4M} \in S_1$, where $\beta_1 \in \{ 0, 1 \}$. Bob's goal is to advance the Alice corresponding to $m_0$ to Stage 3 with the bit $\beta_0 = 1 - \beta_1$.
            
            Since it is now the beginning of a megablock, the value of the relevant counter (either $\cnt_0$ or $\knt_0$) has been reset to $0$. Let $\counter_0 = \cnt_0$ if $\knt_0 = -1$, and otherwise let $\counter_0 = \knt_0$. 
            
        
        
            In either case, the strategy is the same: In each megablock, Bob performs a similiar strategy as in Phase 1 to increment Alice's counter $\counter_0$ to a value $j$, and then send her $\bar{0}$'s for the rest of the megablock to advance her to Stage $3$ with the bit $\beta_0 = 1-\beta_1$. The value $j$ is defined as follows: $j = 1-\beta_1$ if $\knt_0 = -1$ and $j = \beta_1$ if $\knt_0 \in \{ 0, 1 \}$. 
            
            \begin{itemize}
            \item 
                If $j = 0$, Bob sends $\bar{0}$ for the rest of the protocol.
                
                He continues list-decoding Alice's messages with $< \frac34-\frac32\epsilon$ erasures to $\{ m_0, m_1 \}$. If $S_b \cap \{ m_0, m_1 \} = \emptyset$ for some $b \in \{ 0, 1 \}$, he sets $\hat{x} \gets \hat{x}_{1-b}$.
            \item
                If $j = 1$, Bob does the following: he reads Alice's message $m$ and sends a message based on the following cases.
                \begin{caseof}
                \case{$m$ has at least $\frac34 - \frac32\epsilon$ $\perp$'s.}{
                    Bob sends the same message he sent previously, unless it's the start of a block, in which case he sends $\bar{1}$.
                }
                \case{Bob list decodes to $\le 2$ messages $\{ m_0, m_1 \}$, and $S_b \cap \{ m_0, m_1 \} = \emptyset$ for some $b \in \{ 0, 1 \}$. Note that $S_1 = \{ \beta_1^{4M} \}$.}{
                    Bob sets $\hat{x}\gets\hat{x}_{1-b}$.
                }
                \end{caseof}
                
                For the remaining cases, Bob can decode Alice's message into two options: $\ECC(x_0, \cnt_0, \sent_0, \rec_0, \knt_0, \stg_0 = \false)$ and $\beta_1^{4M}$. 
                
                \begin{caseof}
                \setcounter{casenum}{3}
                
                \case{$\counter_0 = 0$.}{
                    Bob sends $\bar{1}$.
                }
                
                \case{$\counter_0 = 1$.}{
                    Bob sends $\bar{0}$ for the rest of the megablock. 
                }
                \end{caseof}
            \end{itemize}
                
            At the end of the protocol, let $b$ be the most recent bit received by Bob. He sets $\hat{x} \gets \hat{x}_1$ if $b = \beta_1$ and $\hat{x} \gets \hat{x}_0$ otherwise. 
        \end{enumerate}
}

\subsection{Analysis} \label{sec:3/5-analysis}

\subsubsection{Correctness Lemmas}

We begin with a lemma that explains the correctness of our protocol. There are two parts: the first is that $S_0$ and $S_1$ can never overlap, that is, if the adversary is confusing Bob between two values of $x$, it is never possible that Alice's messages in the two worlds are the same. The second justifies an assumption we made in the protocol description, which is that if Bob 2-decodes Alice's message, then the two Alices cannot both be in Stage 2. We remind the reader that once Bob has set $\hat{x}$, he no longer updates (or initializes) the sets $S_0, S_1$.

\begin{lemma} \label{lemma:3/5-S0S1}
    At any point in Protocol~\ref{prot:3/5} after $S_0, S_1$ are initialized, 
    \begin{enumerate}[label=(\roman*)]
        \item \label{item:S0S1-different} $S_0 \cap S_1 = \emptyset$.
        \item \label{item:S0S1-knt01} At most one of $S_0, S_1$ contains a message of the form $\ECC(x, \cnt, \sent, \rec, \knt, \stg)$ with $\knt\in \{0,1\}$.
    \end{enumerate}
\end{lemma}

We delay the proof of Lemma~\ref{lemma:3/5-S0S1} to the end of this section and first state a couple corollaries of the fact that $S_0 \cap S_1 = \emptyset$. 







\begin{corollary} \label{cor:unique}
    If it is ever the case that Bob uniquely decodes Alice's message, then Bob outputs $x$ correctly. In particular, in order for Bob to output incorrectly, the adversary must erase at least half of each of Alice's messages.
\end{corollary}

\begin{proof}
    This is true the first time Alice's message to Bob has fewer than $\frac34 - \frac32\epsilon$ erasures. 
    
    Otherwise, consider the first message $m$ from Alice that Bob uniquely decodes to $m'$. By Lemma~\ref{lemma:3/5-S0S1}, $S_0 \cap S_1 = \emptyset$ so $m'$ can only belong to one of $S_0$ and $S_1$. Since Alice's true message must always belong to the set $S_b$ if her input is $\hat{x}_b$, this allows Bob to output $x$ correctly by setting $\hat{x} \gets \hat{x}_b$ where $b$ is such that $m' \in S_b$. This is in fact what he does in the protocol.
\end{proof}

\begin{corollary} \label{cor:inconsistent}
    After $\hat{x}_0, \hat{x}_1, S_0, S_1$ are initialized, if Bob ever $2$-decodes Alice's message to $\{ m_0, m_1 \}$ such that $S_b \cap \{ m_0, m_1 \} = \emptyset$ for some $b$, then Bob outputs $x$ correctly.
\end{corollary}

\begin{proof}
    Alice's true message must always belong to the set $S_b$ if her input is $\hat{x}_b$. Then, if $S_b \cap \{ m_0, m_1 \} = \emptyset$ for some $b$, it must be the case that $x = \hat{x}_{1-b}$. In the protocol, Bob in fact sets $\hat{x} \gets \hat{x}_{1-b}$ if this is the case.
\end{proof}

\begin{proof}[Proof of Lemma~\ref{lemma:3/5-S0S1}]
    
    Assume for the sake of contradiction there exists an execution of the protocol in which at some point after $S_0, S_1$ are initialized, one of the conditions~\ref{item:S0S1-different},\ref{item:S0S1-knt01} is violated. Let $V$ be the first chunk after which this is true, and let $W \le V$ be the last chunk in which Alice's message to Bob was $< (\frac34 - \frac32\epsilon)$-erased, so that Bob list-decoded Alice's message to two possibilities $m_0$ and $m_1$, where not both $m_0$ and $m_1$ are of the form $\beta^{4M}$. It must be that Bob has not set $\hat{x}$ as of chunk $V$, since as soon as he has set $\hat{x}$ he no longer updates (or initializes) $S_0, S_1$. It follows that either Alice's message was $< (\frac34 - \frac32\epsilon)$-erased so that Bob list-decoded Alice's message to two possibilities (with $\hat{x}_0 \not= \hat{x}_1$ and $\knt_0 = \knt_1 = -1$) for the first time in chunk $W$, or $m_0 \in S_0$ and $m_1 \in S_1$, where $S_0, S_1$ are Bob's sets after chunk $W-1$ that satisfy properties~\ref{item:S0S1-different} and~\ref{item:S0S1-knt01}. In either case, it holds that $m_0 \not= m_1$ and not both $m_0, m_1$ have $\knt_b \in \{ 0, 1 \}$. 
    
    We will show that if $m_0 \not= m_1$ and not both have $\knt_b \in \{ 0, 1 \}$, then properties~\ref{item:S0S1-different} and~\ref{item:S0S1-knt01} hold for sets $S_0$ and $S_1$ after chunk $V$ as well. We do so by casework on $m_0, m_1$.
    
    \begin{caseofp}
    \begin{mdframed}[topline=false,rightline=false,bottomline=false]
    \casep{$m_b = \ECC(\hat{x}_b, \cnt_b, \sent_b, \rec_b, \knt_b, \stg_b) ~\forall b \in \{ 0, 1 \}$ and $\knt_0=\knt_1 = -1$.}{
        
        \begin{subcaseofp}
            \subcasep{$\cnt_0 = \cnt_1 < 2i$ and $\rec_0 = \rec_1 = \true$. Bob has sent $\bar{0}$'s for the rest of the block since chunk $W$ followed by $\bar{1}$'s in subsequent blocks.}{
                Note that in neither world can Alice advance: if she hears a $0$, she can at most confirm her current value of $\cnt$. Then, all values in $S_b$ after chunk $V$ must still have $\knt_b = -1$ and $x_b = \hat{x}_b$, so that $S_0 \cap S_1 = \emptyset$ and no elements of either $S_b$ have $\knt \in \{ 0, 1 \}$. 
            }
            \subcasep{$\cnt_0 = \cnt_1 < 2i$ and $\rec_0 \not= \rec_1$. Bob has sent $\bar{0}$'s for the rest of the block since chunk $W$ followed by $\bar{1}$'s in subsequent blocks.}{
                W.l.o.g. suppose $\rec_0 = \true$ and $\rec_1 = \false$.
                Note that since $\rec_0 = \true$ then $\sent_0 = \false$. Then $S_0$ consists only of messages with $\knt = -1$ and $x_0 = \hat{x}_0$, as the corresponding Alice could only have confirmed her counter or incremented further. On the other hand, the set $S_1$ can only consist of states of the form $\beta^{4M}$ or $\ECC(\hat{x}_1, \cnt_1, \sent_1, \rec_1, \knt_1, \stg_1)$ with $\hat{x}_1 \not= \hat{x}_0$. Thus, $S_0 \cap S_1 = \emptyset$ and $S_0$ contains no messages with $\knt \in \{ 0, 1 \}$.
            }
            \subcasep{$\cnt_0 = \cnt_1 < 2i$ and $\rec_0 = \rec_1 = \false$. Bob has sent $\bar{1}$'s for the rest of the block since chunk $W$ and in subsequent blocks.}{
                Neither state could have advanced, since no $0$'s were sent between chunks $W$ and $V$. This means that $S_b$ consists of only elements with $x_b = \hat{x}_b$ and $\knt = -1$, where $\hat{x}_0 \not= \hat{x}_1$, so $S_0 \cap S_1 = \emptyset$ and neither $S_b$ contains a message with $\knt \in \{ 0, 1 \}$. 
            }
            \subcasep{$\cnt_0 \not= \cnt_1$ or $\cnt_0 = \cnt_1 = 2i$. Bob has sent $\bar{0}$'s for the rest of the megablock since chunk $W$, and $\bar{1}$'s in subsequent megablocks.}{
                We are only concerned with the elements of $S_0, S_1$ that do not have $\knt = -1$. That is, the interesting elements are those that Alice would send had she heard a $\bar{0}$ from Bob in the megablock containing chunk $W$. 
                
                First, at most one of the sets $S_0, S_1$ can contain an element with $\knt \in \{ 0, 1 \}$. This is because Bob begins sending $\bar{0}$'s the \emph{first time} he sees either $\cnt_0 \not= \cnt_1$ or $\cnt_0 = \cnt_1 = 2i$, meaning that either $\cnt_0 = \cnt_1 \pm 1$ or $\cnt_0 = \cnt_1 = 2i$, where $\hat{x}_0[i] \not= \hat{x}_1[i]$. It's thus not possible that both Alices satisfy the condition for advancing to Stage 2, namely, $\cnt_b~\text{mod}~2 = 0$ and $\hat{x}_b[\cnt_b/2] = 1$. 
                
                Second, the sets $S_0, S_1$ cannot contain the same element $\beta^{4M}$. This is because Alice only advances to Stage 3 if $\cnt~\text{mod}~2 = 0$ and $x[\cnt/2] = 0$ (with $\beta = 0$), or $\cnt~\text{mod}~2 = 1$ (with $\beta = 1$). Since either $\cnt_0 = \cnt_1 \pm 1$ or $\cnt_0 = \cnt_1 = 2i$, where $\hat{x}_0[i] \not= \hat{x}_1[i]$, the two Alices cannot both have $\cnt~\text{mod}~2 = 0$ and $x[\cnt/2] = 0$ or both have $\cnt~\text{mod}~2 = 1$. 
            }
        \end{subcaseofp}
    }
    \casep{One of $m_0, m_1$ is of the form $\beta^{4M}$.}{
        W.l.o.g. suppose $m_1 = \beta_1^{4M}$ and $m_0 = \ECC(\hat{x}_0, \cnt_0, \sent_0, \rec_0, \knt_0, \stg_0)$. Then $S_1 = \{ \beta_1^{4M} \}$. We will show that $\beta_1^{4M} \not\in S_0$.
        
        If Bob has only sent $\bar{1}$'s to Alice since chunk $W$, then $S_0$ contains no elements of the form $\beta^{4M}$ as Alice cannot advance without receiving a $\bar{0}$.
        
        Bob has only sent a positive number of $\bar{0}$'s since chunk $W$ if either $j = 1$ and $\counter_0 = 1$ (in which case Bob sends $\bar{0}$'s for the rest of the megablock containing $W$) or $j = 0$ and $V$ is in a different megablock than $W$ (in which case Bob sends $\bar{0}$ only in the megablocks after chunk $W$). Recall that $j$ is chosen such that Alice can only advance to Stage 3 with the bit $1 - \beta_1$ if $\counter_0 = j$. Furthermore, Bob only sends $\bar{0}$'s when he knows that Alice's value of $\counter_0$ is equal to $j$. Thus, $\beta_1^{4M} \not\in S_0$.
    }
    \casep{One of $m_0, m_1$ has $\knt = -1$ while the other has $\knt \in \{ 0, 1 \}$.}{
        W.l.o.g. suppose $m_0$ has $\knt_0 = -1$ and $m_1$ has $\knt_1 \in \{ 0, 1 \}$. Recall also that in order for Alice to be in Stage 2, it must be the case that $\cnt_1~\text{mod}~2 = 0$ and $\hat{x}_1[\cnt_1/2] = 1$.
        
        It is impossible that $S_0$ contains an element with $\knt \in \{ 0, 1 \}$, as Alice doesn't advance unless she receives a $\bar{0}$ and Bob only sends $\bar{0}$'s when he's sure that Alice's $\counter_0 = j$, for which Alice would advance directly to Stage 3. Therefore, $S_0$ and $S_1$ do not both have an element with $\knt \in \{ 0, 1 \}$.
        
        We next show that $S_0$ and $S_1$ cannot contain the same element $\beta^{4M}$. If $V$ is in the same megablock as the first time Bob list-decoded to two messages with $\knt_0 = -1$ and $\knt_1 \in \{ 0, 1 \}$, then Bob has only sent $\bar{1}$'s since then and in particular since chunk $W$. Alice cannot advance without receiving a $\bar{0}$, so that $S_0$ and $S_1$ in fact contain no elements of the form $\beta^{4M}$. Otherwise, Bob has only sent $\bar{0}$'s after that first megablock. Recall since Alice resets $\counter_b$ to $0$ at the beginning of megablocks, she could only have received a $\bar{0}$ if $\counter_b = 0$. Then $S_0$ can contain the element $0^{4M}$ (but not $1^{4M}$) and $S_1$ can contain the element $1^{4M}$ (but not $0^{4M}$). Thus $S_0 \cap S_1 = \emptyset$.
    }
    \end{mdframed}
    \end{caseofp}
\end{proof}

\subsubsection{Main Theorem: $3/5$ Erasure Resilience}

We now prove our main theorem that Protocol~\ref{prot:3/5} is resilient to $\frac35 - O(\epsilon)$ adversarial erasures. First, we establish a list of lemmas that we will use in the proof of Theorem~\ref{thm:3/5}.

\begin{lemma} \label{lemma:seen-phase3}
    Let $R$ be the megablock in which Alice advances to Stage 3, or $R = A$ if she never advances to Stage 3. If Bob is in Phase 2 or 3 at the end of the protocol, and if he receives any bit sent after megablock $R$, then Bob outputs $x$ correctly.
\end{lemma}

\begin{proof}
    If Bob is in Phase 2 at the end of the protocol, this means that at some point the two list-decoded states are such that $\knt_0 = -1$ and $\knt_1 \in \{ 0, 1 \}$. When Bob transitions to Phase 2 at the start of a megablock, it must still hold that $\knt_0 = -1$ for all $m_0 \in S_0$ and $\knt_1 \in \{ 0, 1 \}$ for all $m_1 \in S_1$, since Bob only sent $\bar{1}$ for the rest of the previous megablock so that Alice could not further advance. In other words, if the real Alice has input $x = \hat{x}_0$, she must still be in Stage 1, and if she has $x =\hat{x}_1$, she must be in Stage 2. Bob then sends $\bar{0}$ for the rest of the protocol. If the real Alice had $x = \hat{x}_0$ and were in Stage 1, when she receives a bit from Bob, she advances directly to Stage 3 with the bit $B = 0$. If she had $x = \hat{x}_1$ and were in Stage 2, when she receives a bit from Bob, she advances to Stage 3 with $B = 1$. Thus, if Alice advanced and any subsequent bit is received by Bob, he outputs $x$ correctly.
    
    If Bob is in Phase 3 at the end of the protocol, this means that he at some point list-decoded Alice's message to two messages, one of the form $B^{4M}$ and the other of the form $\ECC(\hat{x}_b, \cnt_b, \sent_b, \rec_b, \knt_b, \stg_b)$. If Alice's real input were $\hat{x}_b$, note that by design, she can only advance to Stage 3 with the bit $1-B$. Thus, if Alice advanced to Stage 3 and at least one subsequent bit was received by Bob, he outputs $x$ correctly.
\end{proof}

\begin{lemma} \label{lemma:phase1-2}
    If Bob begins a megablock in Phase 1 and does not transition to Phase 2 or 3 at the end of the megablock, nor has he set $\hat{x}$ by the end of the megablock, the adversary must have erased at least $\frac35-\frac{18}5\epsilon$ of the megablock.
\end{lemma}

\begin{proof}
    Let $E$ be the last block where fewer than $\frac34 - \frac32\epsilon$ of Alice's bits were erased. Note that if no such block exists, then the adversary trivially erased at least
    \[
        \frac{\left( \frac34 - \frac32\epsilon \right) \cdot B \cdot C \cdot 4M}{B \cdot C \cdot 5M} = \frac35 - \frac65\epsilon > \frac35 - \frac{18}{5}\epsilon
    \]
    of the megablock, so we may assume such a block exists.
    Within this block, there is a chunk in which Alice's message is $< (\frac34 - \frac32\epsilon)$-erased, so that Bob list-decoded her message to two messages, both of which must have $\knt = -1$ as Bob does not transition to Phase 2 or 3 at the end of the megablock. This means that as of right before block $E$, both the real and the fake Alice are still in Stage 1.
    
    Since Alice never increments past $2i$, before block $E$, there were at most $2i \le 2n$ blocks in which Alice heard at least one $\bar{0}$ after hearing some nonzero number of $\bar{1}$'s. In the other $\ge (E-1) - 2n$ blocks, Alice heard only $\bar{1}$'s and erased messages. In such blocks, let $V$ be the first chunk in which Bob decodes Alice's message to two possibilities, one in $S_0$ and one in $S_1$ such that $\rec_0 \vee \rec_1 = \true$, so that he sends $\bar{1}$'s before chunk $V$ and $\bar{0}$'s in or after chunk $V$. Since Alice only hears $\bar{1}$'s, she does not receive any of Bob's messages in or after chunk $V$. Furthermore, since chunk $V$ was the first chunk in which Bob list-decodes Alice's message to two messages such that at least one value of $\rec$ is $\true$, it must be the case that in all but one chunk before $V$, the adversary erased either $\frac34 - \frac32\epsilon$ of Alice's message or all of Bob's and $\frac12 - \epsilon$ of Alice's. To summarize, in each of the $\ge (E-1) - 2n$ blocks, at least 
    \begin{align*}
        &\frac{(V-2) \cdot \min \left\{ (\frac34 - \frac32\epsilon) \cdot 4M, (\frac12 - \epsilon) \cdot 4M + M \right\} + (C-V+1) \cdot \left( (\frac12 - \epsilon) \cdot 4M + M \right)}{C \cdot 5M} \\
        \ge& \left( \frac35 - \frac65\epsilon \right) \cdot \frac{C-1}{C} \\
        \ge& \left( \frac35 - \frac65\epsilon \right) \cdot \left( 1 - \epsilon \right) \\
        \ge& \frac35 - \frac95\epsilon
    \end{align*}
    of the rounds were erased, where we used that $C = \lceil \frac1\epsilon \rceil$.
    Combining this with the fact that after block E, $\frac34 - \frac32\epsilon$ of each of Alice's messages is erased, we get that 
            \begin{align*}
                &\ge \frac{(E - 1 - 2n) \cdot \left( \frac35 - \frac95\epsilon \right) \cdot C \cdot 5M + (B - E) \cdot \left( \frac34 - \frac32\epsilon \right) \cdot C \cdot 4M}{B \cdot C \cdot 5M} \\
                &= \frac{(\frac35 - \frac65\epsilon) \cdot B - \frac35\epsilon E - ( \frac35 - \frac95\epsilon) \cdot (2n+1)}{B} \\
                &\ge \left( \frac35 - \frac95\epsilon \right) \left( 1 - \frac{2n+1}{B} \right) \\
                &\ge \left( \frac35 - \frac95\epsilon \right) \left( 1 - 3\epsilon \right) \\
                &\ge \frac35 - \frac{18}{5}\epsilon
            \end{align*}
            of the megablock must have been erased, where we used that $B = \lceil \frac{n}{\epsilon} \rceil$.
\end{proof}

\begin{lemma} \label{lemma:phase23-stage3}
    If Bob begins a megablock in Phase 2 or 3, and Alice is not in Stage 3 nor has Bob set $\hat{x}$ by the end of the megablock, the adversary must have corrupted at least $\frac35-\frac{12}5\epsilon$ of the megablock.
\end{lemma}

\begin{proof}
    If Bob begins the megablock in Phase 2, he sends $\bar{0}$'s for the entire megablock. Note that Alice has reset the value of her $\counter$ to $0$, so that upon receiving a $\bar{0}$ from Bob she advances straight to Stage 3. Thus, in order for Alice not to advance to Stage 3 in this megablock, the adversary must erase all of Bob's messages. They must also erase at least $\frac12-\epsilon$ of each of Alice's messages, or by Lemma~\ref{cor:unique} Bob correctly sets $\hat{x}$ to $x$.
    
    Thus, the adversary must corrupt a total of
    \begin{align*}
        \ge \frac{\left( \frac12 - \epsilon \right) \cdot B \cdot C \cdot 4M + B \cdot C \cdot M}{B \cdot C \cdot 5M} 
        \ge \frac35 - \frac45\epsilon
        > \frac35 - \frac{12}{5}\epsilon
    \end{align*}
    of the megablock.
            
    If Bob begins the megablock in Phase 3 and Alice does not advance to Stage 3, there are $j \le 1$ blocks in which Alice hears a $\bar{0}$. In the other $\ge B - 1$ blocks, Alice hears only $1$'s. Then, in each of these $\ge B-1$ blocks, as in the proof of Lemma~\ref{lemma:phase1-2}, $\ge \frac35 - \frac95\epsilon$ of the rounds must have been erased.
    
    As such, we get that
    \begin{align*}
        \ge& \frac{(B-1)\cdot (\frac35 - \frac95\epsilon)\cdot C\cdot 5M}{B\cdot C \cdot 5M} \\
        \ge& \left(\frac35 - \frac95\epsilon\right) \cdot \left( 1 - \frac1B \right) \\
        \ge& \left(\frac35 - \frac95\epsilon\right) \cdot \left( 1 - \epsilon \right) \\
        \ge& \frac35 - \frac{12}5\epsilon
    \end{align*}
    of the megablock must have been erased.
\end{proof}


    

We are now ready to prove that Protocol~\ref{prot:3/5} is resilient to $\frac35-\frac{24}5\epsilon$ erasures.

\begin{theorem} \label{thm:3/5}
    Protocol~\ref{prot:3/5} is resilient to a $\frac35-\frac{24}{5}\epsilon$ fraction of erasures. For an input of size $n$, the total communication is $O_\epsilon(n^2)$. Alice and Bob run in time $\poly_\epsilon(n)$.
\end{theorem}

\begin{proof}
    The communication complexity is $A\cdot B\cdot C \cdot 5M = \lceil \frac1\epsilon \rceil \cdot \lceil \frac{n}{\epsilon} \rceil \cdot \lceil \frac{1}{\epsilon} \rceil \cdot O_\epsilon(n) = O_\epsilon(n^2)$. The runtimes of Alice and Bob is governed by the time it takes to list-decode Alice's messages, which is $\poly_\epsilon(n)$ by Theorem~\ref{thm:ECC}. We focus on proving error resilience.
    
    Suppose that the adversary erases some number of rounds such that Bob outputs the incorrect value of $x$ at the end of the protocol. This in particular means that Bob never uniquely decodes Alice's message, nor does he decode to two messages, neither of which are in the set $S_b$ for some $b \in \{ 0, 1 \}$, otherwise by Lemmas~\ref{cor:unique} and~\ref{cor:inconsistent} Bob sets $\hat{x} \gets x$ correctly. We will show that there must have been at least $\frac35 - \frac{24}{5}\epsilon$ erasures. 
    
    Let $R \in [A]$ be the megablock in which Alice advances to Stage 3; if she never advances to Stage 3, let $R = A$. If Bob ends the protocol in Phase 1, it must be the case that each of Alice's messages after the $R$'th megablock were $(\frac34 - \frac32\epsilon)$-erased, otherwise Bob would transition from Phase 1 to Phase 3 upon seeing one world with $\knt = \infty$. If Bob is in Phase 2 or 3 at the end of the protocol, Bob couldn't have seen any bit from Alice sent after the $R$'th megablock, otherwise by Lemma~\ref{lemma:seen-phase3} he correctly outputs $x$. In either case, it is the case that all messages from Alice after the $R$'th megablock are at least $(\frac34 - \frac32\epsilon)$-erased.
    
    We claim that there is at most one megablock before megablock $R$ with fewer than $\frac35-\frac{18}5$ erasures. In the first such block, Bob must transition from Phase 1 to Phase 2 or 3 by Lemma~\ref{lemma:phase1-2}. In the second block with fewer than $\frac35 - \frac{18}{5}\epsilon$ erasures, by Lemma~\ref{lemma:phase23-stage3} Alice must have advanced to Stage 3 by its end. Thus, this second megablock cannot be before megablock $R$, as Alice advances to Stage 3 in megablock $R$ (or not at all). Therefore, in each of $R-2$ of the first $R-1$ megablocks, at least $\frac35-\frac{18}5$ of the megablock is erased. Combining this with the fact that in each of the final $A-R$ megablocks, Alice’s messages are at least $(\frac34-\frac32\epsilon)$-erased, we get that
    \begin{align*}
        &\ge \frac{(R-2) \cdot (\frac35 - \frac{18}{5}\epsilon) \cdot B \cdot C \cdot 5M + (A-R)  \cdot (\frac34 - \frac32\epsilon) \cdot B \cdot C \cdot 4M}{A \cdot B \cdot C \cdot 5M} \\
        &= \frac{(\frac35 - \frac65\epsilon)A - \frac{12}{5}\epsilon R - (\frac65 - \frac{36}{5}\epsilon)}{A} \\
        &\ge \frac35 - \frac65\epsilon - \frac{12}{5}\epsilon - \frac{6/5}{A} \\
        &\ge \frac35 - \frac{6}{5}\epsilon - \frac{6}{5} \epsilon \\
        &= \frac35 - \frac{24}{5}\epsilon
    \end{align*}
    of the rounds in the protocol must have been erased, where we use that $R \le A$ and $A = \lceil \frac1\epsilon \rceil$.
\end{proof}

\section{Impossibility Bounds}

\subsection{Binary Erasure Channel} \label{sec:2/3}


\begin{theorem} \label{thm:2/3}
    No $\iECC$ is resilient to a $\frac23$ fraction of erasures with probability greater than $\frac12$.
\end{theorem}

\begin{proof}
    Consider the frequency with which Bob speaks. If he speaks $r \le \frac13$ of the time, consider the following attack: The adversary erases all of Bob's messages. Then, Alice's messages to Bob reduce to an error correcting code, which has relative distance at most $\frac12$. Thus, the adversary can erase half of Alice's messages such that Bob cannot distinguish between two of Alice's inputs. This attack uses $r + \frac{1-r}{2} = \frac{1+r}{2} \le \frac23$ erasures.
    
    Alternatively, if Bob speaks $r > \frac13$ of the time, the adversary can simply erase all of Alice's messages to Bob. Then, Bob receives no information about Alice's input. This attack requires $< \frac23$ erasures. 
    
    Thus, there is always an attack using $\le \frac23$ erasures for which Bob cannot correctly output Alice's input with probability greater than $\frac12$. 
\end{proof}



\subsection{Binary Bit Flip Channel} \label{sec:2/7}

We also propose the problem of constructing $\iECC$s in the standard bit flip corruption model. It is trivial to construct an $\iECC$ resilient to $\frac14$ adversarial bit flips: simply have Alice send $\ECC(x)$, where $\ECC$ is a standard error correcting code of relative distance $\frac12$. The following is a corollary of Theorem~\ref{thm:ECC-plain}.

\begin{proposition}
    For any $\epsilon>0$, there exists an $\iECC$ resilient to $\frac14-\epsilon$ adversarial bit flips such that the communication complexity for inputs of size $n$ is $O_\epsilon(n)$. 
\end{proposition}

    
    

Without interaction, one cannot achieve error resilience greater than $\frac14$. Unfortunately, we do not know of any protocols that use interaction to achieve a higher error resilience. We leave constructing one an open problem.

Instead, we prove the following impossibility bound on the error resilience of any $\iECC$ in the bit flip setting. 

\begin{theorem} \label{thm:2/7}
    No $\iECC$ is resilient to more than $\frac27$ errors with probability greater than $\frac12$.
\end{theorem}

\noindent
In our proof, we use the following classical result.

\begin{theorem}\cite{SpencerW92} \label{thm:three-liars}
    No $\iECC$ is resilient to an attack consisting of corrupting $\frac13$ of Alice's messages and none of Bob's messages with probability greater than $\frac12$.
\end{theorem}

\begin{proof}[Proof of Theorem~\ref{thm:2/7}]
    Let us first assume that Alice talks at most $\frac67$ of the time. By Theorem~\ref{thm:three-liars}, even if none of Bob's messages are corrupted, the adversary has a strategy to confuse Bob between two inputs by
    corrupting $\frac13$ of the bits Alice sends. This attack requires results at most $\frac13\cdot \frac67= \frac27$ corruptions.

    Now, let us assume that Alice talks at least $\frac67$ of the time. Suppose that Alice talks in $A$ rounds and Bob talks in $B$ rounds. We can divide the protocol into $C$ chunks in which Alice talks contiguously for some number of rounds then Bob speaks contiguously for some number of rounds (the number of rounds within each chunk does not have to be the same). Let Alice's possible inputs be $x_1, \dots, x_N$.  For $k\in \{1\dots C\}$, we define $A_k(x_i)$ and $B_k(x_i,x_j)$ and construct $R_k(x_i,x_j)$ and $S_k$ as follows:
    \begin{itemize}
    \item 
        For each $i \in [N]$, let $A_k(x_i)$ be what Alice would send in the $k$'th chunk if she has input $x_i$ and she's seen the $k-1$ messages $S_1, \dots, S_{k-1}$ from Bob. (If $k = 1$ then $A_k(x_i)$ is simply what Alice would send in the first chunk if her input were $x_i$). 
    \item 
        For each pair $i\not=j$, let $R_k(x_i,x_j)$ be a string that's minimally equidistant from $A_k(x_i)$ and $A_k(x_j)$, satisfying
        \[
            \Delta(R_k(x_i, x_j), A_k(x_i)) = \Delta(R_k(x_i, x_j), A_k(x_j)) = \frac12 \cdot \Delta(A_k(x_i), A_k(x_j)), 
        \]
        This can be constructed by letting $R_1(x_i, x_j)[\ell] = A_1(x_i)[\ell] = A_1(x_j)[\ell]$ when the $\ell$'th bit of $A_1(x_i)$ and $A_1(x_j)$ are the same, and $R_1(x_i, x_j)[\ell] = A_1(x_i)[\ell]$ half the time and $R_1(x_i, x_j)[\ell] = A_1(x_j)$ half the time when $A_1(x_i)[\ell] \not= A_1(x_j)[\ell]$. 
    \item 
        For each $i \not= j$, let $B_k(x_i, x_j)$ be Bob's message if he's seen the messages $R_1(x_i, x_j), \dots, R_k(x_i, x_j)$ from Alice so far.
    \item 
        Define $S_k$ bitwise by setting $S_k[\ell] = \maj \{ B_k(x_i, x_j) \}_{i,j}$.
    \end{itemize}

    Now, for all pairs $i \not= j$, define the transcript $\tau_{ij}$ to be where Bob receives $R_k(x_i, x_j)$ and Alice receives $S_k$ in each chunk $k \in [C]$. We will show that for some $i\not=j$, in both the case Alice has $x_i$ or she has $x_j$ it costs few corruptions for an adversary to corrupt the protocol so that the resulting communication is $\tau_{ij}$. Note that the number of corruptions needed so that the communication is $\tau_{ij}$ is the same whether Alice has $x_i$ or $x_j$, since Bob's message is corrupted identically in both cases and $R_k(x_i, x_j)$ is picked to be equidistant from $A_k(x_i)$ and $A_k(x_j)$.
    
    Consider this attack on the protocol where Alice has $x_i$ and the resulting transcript is $\tau_{ij}$. The total number of corruptions over all (ordered) pairs is:
    \begin{align*}
        \sum_{ij} \sum_k \Big[ \Delta \big( S_k, B_k(x_i, x_j) \big) + \Delta \big( R_k(x_i,x_j), A_k(x_i) \big) \Big] \\
        = \sum_{ijk} \left[ \Delta \big( S_k, B_k(x_i, x_j) \big) + \frac12 \cdot \Delta \big( A_k(x_i), A_k(x_j) \big) \right] \\
        \le \frac12 B \cdot N (N-1) + \frac12 \cdot A \cdot \frac12 N^2 \\
        \le \left( \frac12 B + \frac14 A \right) N^2,
    \end{align*}
    where the inequality $\sum_{ijk} \Delta(S_k, B_k(x_i, x_j)) \le \frac12 B \cdot N(N-1)$ follows from the fact that for each bit $j \in [B]$, $S_k[j]$ is different from at most half of the values of $B_k(x_i,x_j)[j]$, and $\Delta(A_k(x_i), A_k(x_j)) \le A \cdot \frac12 N^2$ follows from the fact that for any position $j \in [A]$, across the $N$ codewords $A_k(x_i)$, there are at most $\frac12 N^2$ pairs of codewords with different values for bit $j$.
    
    Therefore, by the pigeonhole principle, there is some pair $i,j$ for which the attack uses $\le \frac12 B + \frac14 A \le \frac12 \cdot \frac17 + \frac14 \cdot \frac67 = \frac27$ corruptions.
    

    
    
\end{proof}
\section{Acknowledgments}

We would like to thank Venkat Guruswami, Yury Polyanskiy, and Raghuvansh Saxena for enlightning comments and helpful discussions. We would also like to thank Yang P. Liu and Mark Sellke for reading the paper and providing feedback. 

\begin{figure}[h!]
    \centering
    \includegraphics[scale=0.075]{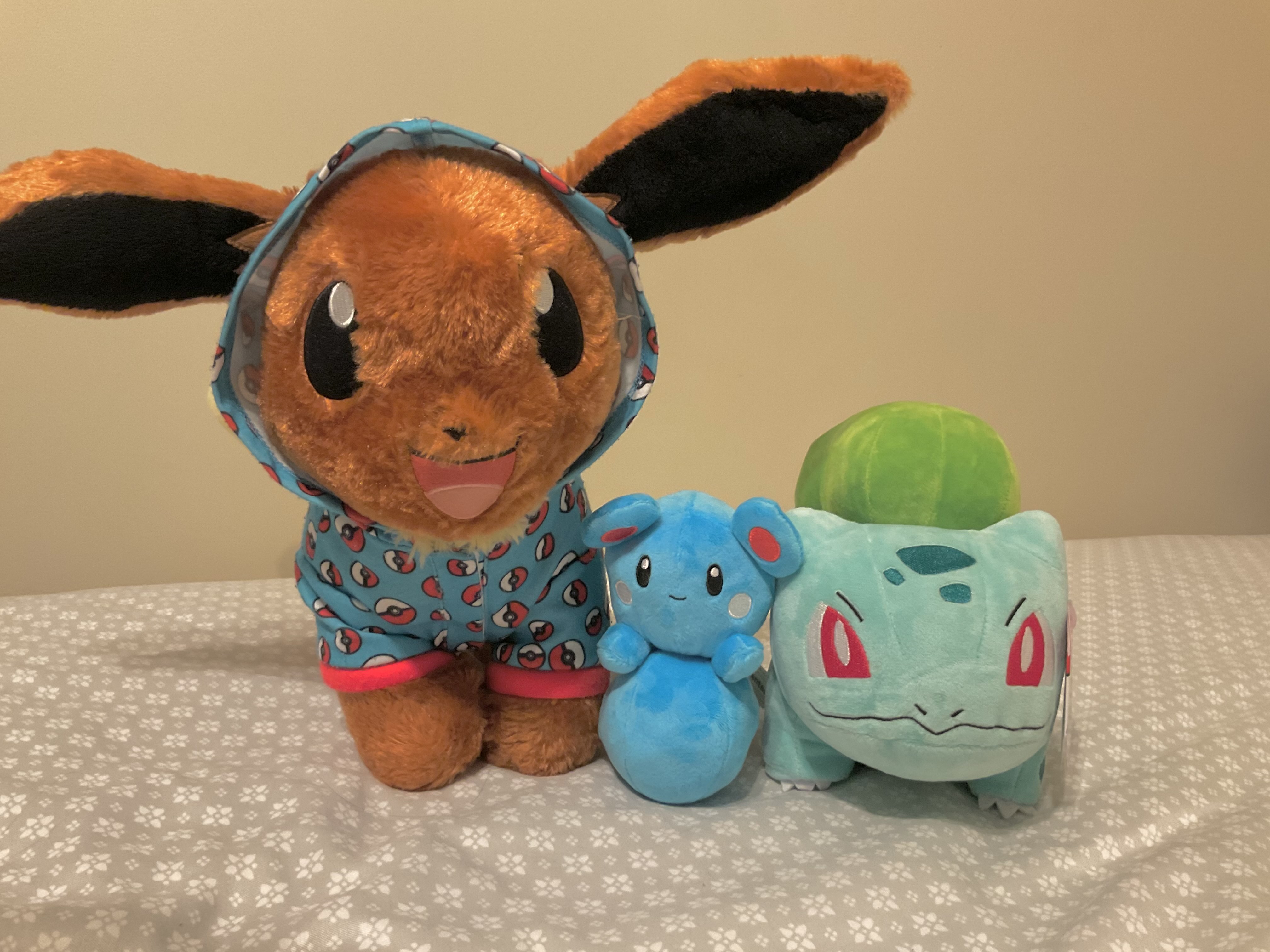}
    \caption{(Left to right) Eevee, Aliceurill, and Bobasaur}
    \label{fig:bobasaur}
\end{figure}

Finally, we want to thank Aliceurill, Bobasaur, and Eevee, without whom this paper would not be possible. As seen in Figure~\ref{fig:bobasaur}, the three remain good friends despite Eevee's occasional adversarial tendencies.

\bibliographystyle{alpha}
\bibliography{refs}

\end{document}